\documentclass[twocolumn]{aastex62}
\let\pwiflocal=\iffalse \let\pwifjournal=\iffalse

\usepackage{amsmath,amssymb}
\usepackage{epsfig}    
\usepackage{graphicx}    
\usepackage{lineno}
\usepackage{natbib}
\usepackage{bigints}
\usepackage[outdir=./]{epstopdf}

\usepackage[T1]{fontenc}
\pwifjournal\else
  \usepackage{microtype}
\fi

\pwifjournal\else
  \makeatletter
  \renewcommand\plotone[1]{%
    \centering \leavevmode \setlength{\plot@width}{0.95\linewidth}
    \includegraphics[width={\eps@scaling\plot@width}]{#1}%
  }%
  \makeatother
\fi

\makeatletter

\newcommand\@simpfx{http://simbad.u-strasbg.fr/simbad/sim-id?Ident=}

\newcommand\MakeObj[4][\@empty]{
  \pwifjournal%
    \expandafter\newcommand\csname pkgwobj@c@#2\endcsname[1]{\protect\object[#4]{##1}}%
  \else%
    \expandafter\newcommand\csname pkgwobj@c@#2\endcsname[1]{\href{\@simpfx #3}{##1}}%
  \fi%
  \expandafter\newcommand\csname pkgwobj@f#2\endcsname{#4}%
  \ifx\@empty#1%
    \expandafter\newcommand\csname pkgwobj@s#2\endcsname{#4}%
  \else%
    \expandafter\newcommand\csname pkgwobj@s#2\endcsname{#1}%
  \fi}%

\newcommand\MakeTrunc[2]{
  \expandafter\newcommand\csname pkgwobj@t#1\endcsname{#2}}%

\newcommand{\obj}[1]{%
  \expandafter\ifx\csname pkgwobj@c@#1\endcsname\relax%
    \textbf{[unknown object!]}%
  \else%
    \csname pkgwobj@c@#1\endcsname{\csname pkgwobj@s#1\endcsname}%
  \fi}
\newcommand{\objf}[1]{%
  \expandafter\ifx\csname pkgwobj@c@#1\endcsname\relax%
    \textbf{[unknown object!]}%
  \else%
    \csname pkgwobj@c@#1\endcsname{\csname pkgwobj@f#1\endcsname}%
  \fi}
\newcommand{\objt}[1]{%
  \expandafter\ifx\csname pkgwobj@c@#1\endcsname\relax%
    \textbf{[unknown object!]}%
  \else%
    \csname pkgwobj@c@#1\endcsname{\csname pkgwobj@t#1\endcsname}%
  \fi}

\makeatother

\pwifjournal\else
  \usepackage{etoolbox}
  \makeatletter
  \patchcmd{\NAT@citex}
    {\@citea\NAT@hyper@{%
       \NAT@nmfmt{\NAT@nm}%
       \hyper@natlinkbreak{\NAT@aysep\NAT@spacechar}{\@citeb\@extra@b@citeb}%
       \NAT@date}}
    {\@citea\NAT@nmfmt{\NAT@nm}%
     \NAT@aysep\NAT@spacechar\NAT@hyper@{\NAT@date}}{}{}
  \patchcmd{\NAT@citex}
    {\@citea\NAT@hyper@{%
       \NAT@nmfmt{\NAT@nm}%
       \hyper@natlinkbreak{\NAT@spacechar\NAT@@open\if*#1*\else#1\NAT@spacechar\fi}%
         {\@citeb\@extra@b@citeb}%
       \NAT@date}}
    {\@citea\NAT@nmfmt{\NAT@nm}%
     \NAT@spacechar\NAT@@open\if*#1*\else#1\NAT@spacechar\fi\NAT@hyper@{\NAT@date}}
    {}{}
  \makeatother
\fi

\providecommand{\adsurl}[1]{\href{#1}{ADS}}

\newcommand{\kepler}{{\it Kepler}}

\newcommand{\um}{$\mu$m}

\newcommand\teff{\ensuremath{T_\text{eff}}}

\usepackage[outdir=./]{epstopdf}
\usepackage{ulem}
\usepackage{mathrsfs}
\usepackage{fontawesome}
\usepackage{comment}

\newcommand{\ktwo}{\textit{K2}}

\newcommand{\mearth}{MEarth}

\newcommand{\spitzer}{\textit{Spitzer}}
\pdfoutput=1

\shorttitle{The young planet, K2-25b} 
\shortauthors{Thao et al.}

\begin{document}

\title{\textbf{Zodiacal Exoplanets in Time (ZEIT) IX: a flat transmission spectrum and a highly eccentric orbit \\ for the young Neptune K2-25b as revealed by \textit{Spitzer}}} 
\correspondingauthor{Pa Chia Thao}
\email{pachia@live.unc.edu; thao22p@mtholyoke.edu} 

\author[0000-0001-5729-6576]{Pa Chia Thao}
\altaffiliation{TAURUS Scholar}
\affiliation{Department of Physics and Astronomy, The University of North Carolina at Chapel Hill, Chapel Hill, NC 27599, USA} 
\affiliation{Department of Astronomy, The University of Texas at Austin, Austin, TX 78712, USA}

\author[0000-0003-3654-1602]{Andrew W. Mann 
}
\affiliation{Department of Physics and Astronomy, The University of North Carolina at Chapel Hill, Chapel Hill, NC 27599, USA}

\author[0000-0002-5099-8185]{Marshall C. Johnson} 
\affiliation{Las Cumbres Observatory, 6740 Cortona Drive, Suite 102, Goleta, CA 93117, USA}

\author[0000-0003-4150-841X]{Elisabeth R. Newton}
\affiliation{Department of Astronomy and Physics, Dartmouth College, Hanover, NH 03755, USA}
\affiliation{Kavli Institute for Astrophysics and Space Research, Massachusetts Institute of Technology, Cambridge, MA 02139, USA}

\author{Xueying Guo}
\affiliation{Kavli Institute for Astrophysics and Space Research, Massachusetts Institute of Technology, Cambridge, MA 02139, USA}

\author[0000-0001-9894-5229]{Isabel J. Kain}
\affiliation{Department of Physics, Northeastern University, Boston, MA 02115, USA}
\affiliation{Kavli Institute for Astrophysics and Space Research, Massachusetts Institute of Technology, Cambridge, MA 02139, USA}

\author[0000-0001-9982-1332]{Aaron C. Rizzuto}
\altaffiliation{51 Peg b Fellow}
\affiliation{Department of Astronomy, The University of Texas at Austin, Austin, TX 78712, USA}

\author[0000-0002-9003-484X]{ David Charbonneau}
\affiliation{Harvard-Smithsonian Center for Astrophysics, Harvard University, Cambridge, MA 02138, USA}

\author[0000-0002-4297-5506]{Paul A. Dalba}
\affiliation{Department of Earth and Planetary Sciences, The University of California Riverside, Riverside, CA 92521, USA}

\author[0000-0002-5258-6846]{Eric Gaidos}
\affiliation{Department of Earth Sciences, The University of Hawaii at M\={a}noa, Honolulu, HI 96822 USA}

\author{Jonathan M. Irwin}
\affiliation{Harvard-Smithsonian Center for Astrophysics, Harvard University, Cambridge, MA 02138, USA}

\author[0000-0001-9811-568X]{Adam L. Kraus}
\affiliation{Department of Astronomy, The University of Texas at Austin, Austin, TX 78712, USA}

\begin{abstract}
Transiting planets in nearby young clusters offer the opportunity to study the atmospheres and dynamics of planets during their formative years. To this end, we focused on K2-25b -- a close-in ($P$=3.48 days), Neptune-sized exoplanet orbiting a M4.5 dwarf in the 650\,Myr Hyades cluster. We combined photometric observations of K2-25 covering a total of 44 transits and spanning $>2$ yr, drawn from a mix of space-based telescopes (\textit{Spitzer Space Telescope} and \textit{K2}) and ground-based facilities (Las Cumbres Observatory Global Telescope network and MEarth). The transit photometry spanned 0.6--4.5$\mu$m, which enabled our study of K2-25b's transmission spectrum. We combined and fit each dataset at a common wavelength within a Markov Chain Monte Carlo framework, yielding consistent planet parameters. The resulting transit depths ruled out a solar-composition atmosphere for K2-25b for the range of expected planetary masses and equilibrium temperature at a $>4\sigma$ confidence level, and are consistent with a flat transmission spectrum. Mass constraints and transit observations at a finer grid of wavelengths (e.g., from the {\it Hubble Space Telescope}) are needed to make more definitive statements about the presence of clouds or an atmosphere of high mean molecular weight. Our precise measurements of K2-25b's transit duration also enabled new constraints on the eccentricity of K2-25's orbit. We find K2-25b's orbit to be eccentric ($e>0.20$) for all reasonable stellar densities and independent of the observation wavelength or instrument. The high eccentricity is suggestive of a complex dynamical history and motivates future searches for additional planets or stellar companions. 
\end{abstract}



\section{Introduction}\label{sec:intro}

A key question of exoplanet research is to understand how planets form and evolve. With the success of the \textit{Kepler} mission \citep{borucki2010kepler} and earlier surveys, our sample of transiting mature exoplanets has expanded in the past decade, allowing us to gain a wealth of information about the late-time configurations of planetary systems and their atmospheres. However, planets are not born in their final states; rather, their dynamical, structural, and atmospheric properties are altered as they interact with their host star, the protoplanetary disk from which they formed, other planets in the system, and their stellar environment \citep[e.g.,][]{Chatterjee2008,Cloutier2013,Kaib:2013}. These processes are the likely strongest during the first 100 Myr after formation; at later times such processes are expected to slow dramatically or enter equilibrium. Comparing the statistical properties of young ($<$\,Gyr) planets to their older ($>$ 1\,Gyr) counterparts is the most direct means to observe evolution, including how planets migrate \citep[e.g.,][]{David2016b, Mann2016b}, lose atmosphere \citep[e.g.,][]{Obermeier2016, rizzuto2018zodiacal}, and cool \citep[e.g.,][]{2015Sci...350...64M}

By studying planets across a wide range of ages, we also measure the timescale of such changes \citep[e.g., ][]{2015MNRAS.452.2127S}, which can be used to test models of their underlying physical drivers. For example, early processes like planetary migration, planet-planet collisions, and dynamical instabilities can heavily influence the orbital eccentricities of planets \citep[e.g.,][]{1996Sci...274..954R, Fabrycky:2007ys, Chatterjee2008, 2008ApJ...678.1407F}. Hence, the distribution of orbital eccentricities for a given planet type, and how that distribution changes with time provides a window into their formation and evolution \citep[e.g.,][]{Dawson:2012fk}.
 
Transit observations with high cadence and precision place constraints on the eccentricity of a transiting planet, provided the density of the host star is known \citep[e.g.,][]{Seager:2003lr, Kipping2012}. While the method is degenerate with the argument of periastron, this technique has been widely successful at constraining the statistical distribution of eccentricities of planets across a wide range of parameter space \citep[e.g.,][]{2008ApJ...678.1407F, Moorhead:2011lr, Dawson:2012fk, Mann2017b}.
 
\cite{Van-Eylen2015} and \citet{vanEylen_eccentricity2019} utilized precise stellar densities from asteroseismology \citep[e.g.,][]{2013ApJ...767..127H,2015MNRAS.452.2127S} to study the eccentricity distribution of small planets from the {\it Kepler} survey. They found that systems with a single detected transiting planet tend to have larger eccentricities than those in multitransiting systems. This is consistent with larger findings from {\it Kepler}, which suggested many systems with a single transiting planet are part of a distinct population with larger mutual inclinations or fewer planets \citep{Ballard2016}, although others have suggested this can be explained by a lower detection efficiency for multiplanet systems \citep{2019MNRAS.483.4479Z}, or a non-Poission planet distribution \citep{Gaidos2016b}. A comparable sample of young planets with eccentricity measurements could demonstrate if this bimodality in the planet population is a consequence of different formation scenarios or different evolutionary paths.
 
Statistical analyses of available masses and radii suggest that most planets larger than 1.6$R_\oplus$ have an envelope with a low mean molecular weight such as of H/He \citep{Rogers2015}. However, the transmission spectra of these planets are generally (though not universally) flat and featureless \citep[e.g., ][]{2010Natur.468..669B,  Berta2012, 2014Natur.505...69K, Crossfield2017}. Since the timescales of atmospheric chemistry is likely to be short, exploring the transmission spectrum of young planets offers the most direct means to test whether the processes of aerosol/cloud formation change over time as a result of changes in the planet's UV irradiation, equilibrium temperature, surface gravity, and other parameters that may change with time.

Here, we focus on the 650\,Myr, Neptune-sized exoplanet, K2-25b \citep{David2016, Mann2016a}, which orbits a cool M4.5 dwarf in the nearby Hyades cluster. Its large transit depth ($\sim$1.1$\%$) and proximity to the Sun (47 parsecs) make it one of the most amenable sub-Neptunes known for transmission spectroscopy \citep{Rodriguez:2017aa}. Compared to its older counterparts around similarly cool stars from \textit{Kepler}, K2-25b has an abnormally large size ($R_{p}=3.45R_{\oplus}$) for its host star mass ($M_{*} = 0.26M_{\odot})$, suggesting that this planet may still be contracting or losing its atmosphere \citep{Mann2016a}.

We combined 20 transits from the discovery \ktwo\ data with 12 transits from the MEarth survey \cite[presented in the companion paper,][]{Kain2019}, 10 new \spitzer\ transit observations, and two new transits from the Las Cumbres Global Observatory Telescope network (Section~\ref{sec:obs}) with the goal of updating the planetary parameters, measuring the eccentricity of the planet, and exploring its atmospheric transmission spectrum. Using the precise parallax from \textit{Gaia}, we updated K2-25's stellar parameters, including the density, as we detail in Section~\ref{sec:params}. We utilized this information in our fit to the transit light curve as described in Section~\ref{sec:transit}. We analyzed our best-fit transit parameters and discuss the atmospheric composition inferred from the transmission spectrum of K2-25b in Section~\ref{sec:results}. In Section~\ref{sec:summary}, we conclude with a brief summary of our results, the need for additional follow-up, and the importance of studying more young planets. 

\section{Observations and Data Reduction}\label{sec:obs}
We collected 44 total transits of K2-25b obtained from 2015 to 2017, taken by \textit{K2}, Las Cumbres Global Observatory Telescope network, the MEarth Project, and the \textit{Spitzer Space Telescope}. The combined data sets span from the visible to the infrared (0.6-4.5$\mu$m). The details of the data are summarized in Table~\ref{tab:obslog}. 

\subsection{\ktwo}
 
We used the \ktwo\ \citep[repurposed \kepler;][]{Howell2014} light curve given in the discovery paper \citep{Mann2016a}, which we briefly describe below. The data covered a total of 20 transits in 71 days from 2015 February 8 to 2015 April 20 (\textit{K2} Campaign 4). 

The extracted \ktwo\ light curves show variations due to telescope drift and pixel-to-pixel variations in the flat field, stellar variability, and individual transits \citep{VanCleve2016}. We fit for all of these effects simultaneously, following the procedure from \citet{Becker2015}. We assigned a single error value to all points, determined using the rms error of out-of-transit points in the detrended light curve. Stellar variability (rotation) and flat-field corrections were both modeled as break-point splines (with break points every 0.2 days and 0.4\arcsec). A transit model was included to avoid biasing the stellar variability and \ktwo\ drift fits. We then used the resulting best-fit model for the stellar and flat-field induced variability to clean the light curve of these effects. Stellar flares, which are seen in the processed light curve, were flagged and manually removed. No obvious flares were observed during a transit. The resulting light curve was used in our Markov Chain Monte Carlo (MCMC) analysis (Section~\ref{sec:transit}). 

\subsection{\spitzer}
We obtained 10 full transits of K2-25b, five in each of 3.6$\mu$m (Channel 1) and 4.5$\mu$m (Channel 2), taken by the Infrared Array Camera \citep[IRAC; ][]{Fazio2004} on the \spitzer\ \textit{Space Telescope} \citep{Werner2004}. Observations were executed over the period of 2016 November 28--2017 May 11 (Program ID: 13037, PI: Mann). We observed each target in the subarray mode, with each image taken in a 2 s exposure of $32\times32$ pixels. Each transit consisted of a $\simeq$30\ minute dither, a $\simeq$110\ minute stare of the full transit, followed by another $\simeq$10\ minute dither\footnote{\href{https://irachpp.spitzer.caltech.edu/}{https://irachpp.spitzer.caltech.edu/}}. The initial dither allows an initial settling time at the new pointing position. For the long stare, we used the peak-up pointing mode, which keeps the star stable on a 0.5$\times$0.5 pixel box region of the IRAC CCD with relatively uniform sensitivity \cite[the sweet spot;][]{Ingalls2012,Ingalls2016}. 

For our analysis, we used the flat-fielded and dark-subtracted basic calibrated data (BCD) images produced by the \spitzer\ pipeline. We tested building our own on-sky dark by median stacking the dithered images, but changes to the light-curve precision were negligible. Therefore, we did not use the on-sky dark in our analysis. We identified cosmic rays by comparing each image to a median stack of 10 consecutive images (five before and five after) and identifying pixels $>6\sigma$ above the stack, then removed the cosmic ray by replacing the pixel value with the mean of all surrounding pixels in both position and time (i.e., using the same pixel and those surrounding it in the preceding and following images). Cosmic rays that overlapped with the photometric aperture (generally a circle with a 3 pixel radius around the centroid) were flagged and these images were not included in our analysis. Less than 1\% of the total images were removed this way. 

We identified the star's position within each image using a flux-weighted centroid with a radius of 2 pixels. We subtracted the background flux, which we estimated from the median of all pixels after masking out a circle with a radius of 4 pixels centered on the object.

\subsubsection{Corrections For Intra-pixel Sensitivity Variations}\label{sec:spitzer}

Due to \spitzer's large intra-pixel sensitivity and its pointing jitter, the measured flux of a source can vary up to 8\%, depending on where it falls on a pixel \citep{Ingalls2012}. Fortunately, years of high-precision observations with \spitzer\ have provided a wealth of methods to extract photometry and correct for model variations in the photometric response \citep[see][for a comparison of methods]{Ingalls2016}. We tested three different methods for decorrelating this instrumental systematic: (1) using a high-resolution pixel variation gain map \citep[PMAP;][]{Ingalls2012}, (2) nearest neighbors \citep[NNBR;][]{Lewis2013}, and (3) pixel-level decorrelation \citep[PLD;][]{Deming2015}. 

\begin{figure*}[htp]
    \includegraphics[width=0.495\linewidth]{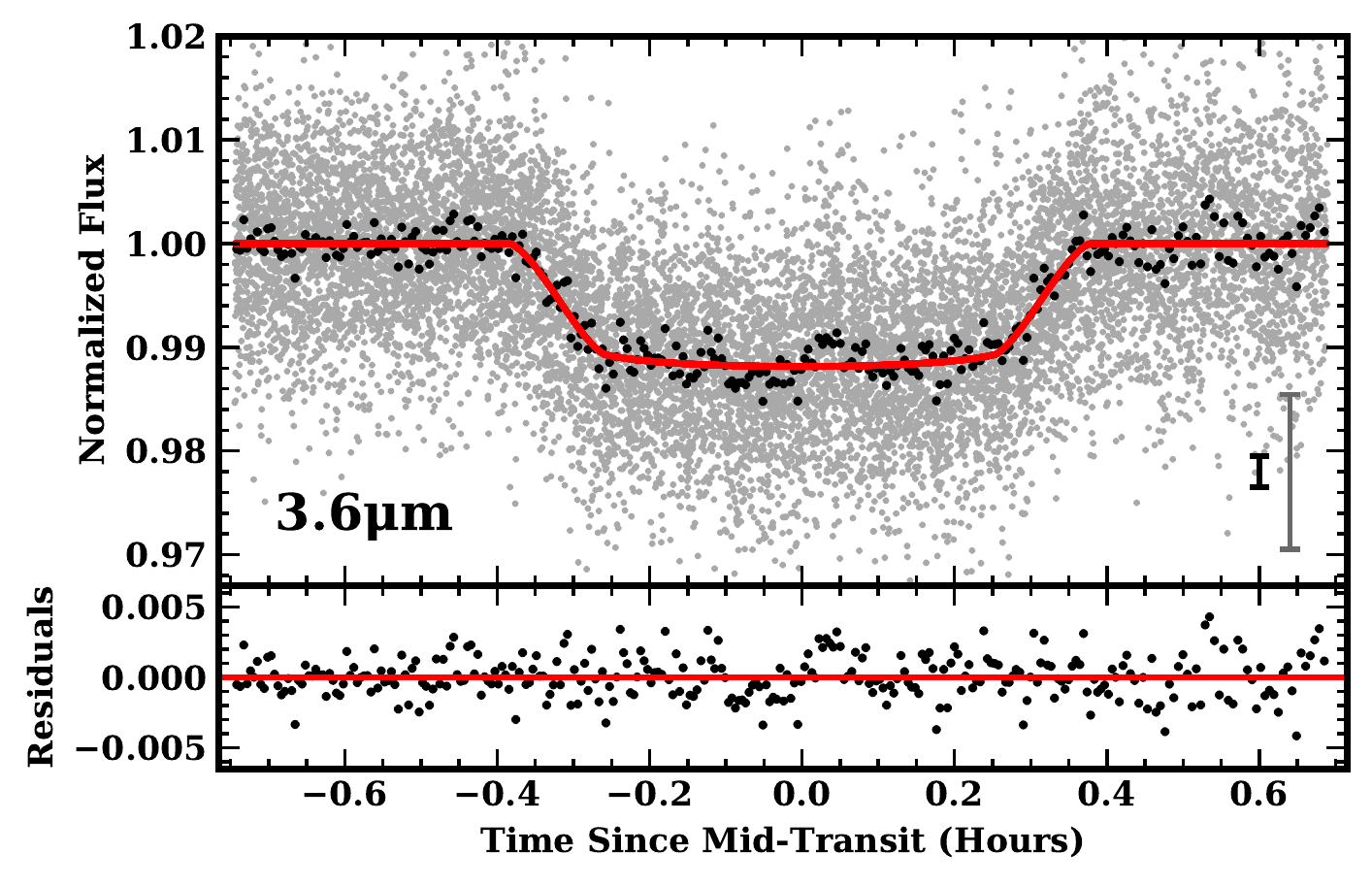}
    \includegraphics[width=0.495\linewidth]{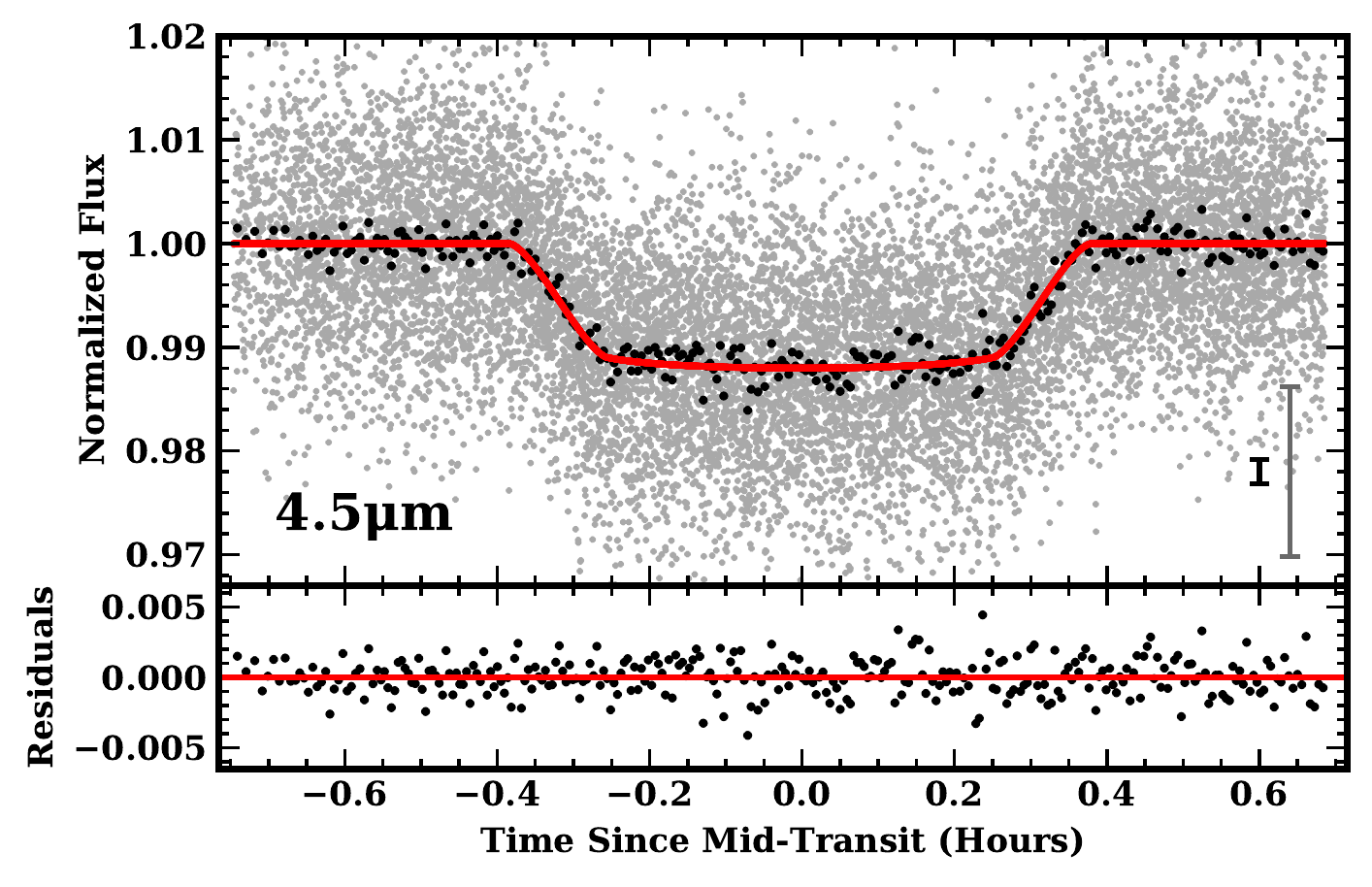}
    \caption{Phase-folded light curve of K2-25's five transits each observed with \textit{Spitzer}/IRAC in 3.6$\mu$m (left) and 4.5$\mu$m (right) with intra-pixel sensitivity correction using PMAP. The individual data points are shown here in gray. The black points correspond to the light curve binned in phase using a median bin size of 80 s. The red line corresponds to the best-fit (highest likelihood) model from our MCMC fit. Typical error bars are derived from scatter in the out-of-transit data points and are shown in the bottom right of the light curve. The black error bar corresponds to the binned data and the gray error bar corresponds to all of the data points. The bottom panel shows the fit residuals using the binned points. \label{fig:spitzerLC}}
\end{figure*}

For PMAP, we used the recommendations from the IRAC program website\footnote{\href{https://irachpp.spitzer.caltech.edu/page/contrib}{https://irachpp.spitzer.caltech.edu/page/contrib}}. We first computed the target point-spread function(PSF) centroid in each image with the IDL routine \texttt{box\_centroider} with a fixed circular aperture radius of 3 pixels, as recommended by IPAC. We used the same circular aperture centered at the source to compute the total flux in each image and passed this along with the $x$ and $y$ positions from the above centroiding routine to \texttt{iracpc\_pmap\_corr} to calculate the corrected flux values. Further details about the photometric gain map are discussed in \citet{Ingalls2012}. The resulting fluxes were fed into our MCMC fitting framework (see Section~\ref{sec:transit}). Figure~\ref{fig:spitzerLC} shows the phase-folded light curves of \spitzer\ Channel 1 and Channel 2 with the corrected fluxes using PMAP.

Each dataset was also corrected using the NNBR technique. We followed the methodology of \citet{Lewis2013}. We took background-subtracted BCD files and calculated the target centroid and total flux in each image using a center of light method, with variable circular aperture radii as described in \citet{Lewis2013}. Each flux value was corrected by linking it to the 50 nearest neighbors weighted by a Gaussian smoothing kernel. We also tested fixed apertures of 2-3 pixels, which yielded consistent transits. Unlike with PMAP, the NNBR correction depends on the transit parameters, so uncorrected fluxes were used as an input to the MCMC fit described in Section~\ref{sec:transit}, and the NNBR correction was computed for each step in the MCMC.

For PLD, we used the procedure given in \citet{Guo2018}, which is based on the procedure outlined in \citet{Deming2015}. This method included binning the data every 64 frames, using all pixels on which the incoming stellar flux fell, formulating their contribution to the total flux over time as eigen-vectors, and setting the weights of those eigen-vectors as free parameters. We included a quadratic time-dependent term to fit the out-of-transit variability. As with NNBR, a PLD correction was computed for each step of the MCMC with a transit model to find the best-fit parameters and their uncertainties as outlined in Section~\ref{sec:transit}.

We fit \spitzer\ light curves using all three detrending techniques as a test; however, we selected which {\textit Spitzer} correction to adopt for our final result based on two criteria: (1) consistency between transit depths at a given wavelength, and (2) overall minimal red noise levels in the fit residuals. While the observed transit depth can vary due to the influence of spots or stellar activity, these effects are expected to be small in the near-infrared, where flares are weaker and the spot contrast is closer to unity. Transit depth variations are more likely to be due to imperfect instrumental corrections.

The PLD fits yielded variation in measured transit depths within a single channel ($\sigma_{Channel1}$ = 0.030; $\sigma_{Channel2}$ = 0.091) larger than the uncertainties on the transit depths themselves ($\sigma_{Channel1}$ = 0.016; $\sigma_{Channel2}$ = 0.018). This was likely because our out-of-transit baseline for some transits was too small to get a reasonable constraint on the intra-pixel sensitivity correction. Due to this, we selected not to use the PLD fit for our analysis. Both PMAP and NNBR yielded transit depth variations that were consistent with the uncertainties on the transit depth.

We quantified how well PMAP, PLD, and NNBR corrections reduced the time-correlated red noise and uncorrelated white noise through the $\beta_{\rm{red}}$ coefficient described in \cite{Gillon2010}: 
\begin{equation}
    \beta_{\rm{red}} = \frac{\sigma_{N}}{\sigma_{1}}
    \sqrt{\frac{N(M-1)}{M}} ,
\end{equation}
where $N$ is the mean number of points in each bin, $M$ is the number of bins, and $\sigma_{N}$ and $\sigma_{1}$ are the standard deviation of the binned and unbinned residuals. This method is commonly used to characterize red noise in \spitzer\ light curves \citep[e.g.,][]{2017AJ....153...22K}. $\beta_{\rm{red}}$ is effectively a measure of how well the scatter in the residuals improve from binning compared to the expectation for perfectly white noise ($\sim\frac{1}{\sqrt{Binsize}}$). To characterize the amount of time-correlated noise at the timescales of ingress and egress ($\sim$ 9 minutes), we used the median $\beta_{\rm{red}}$ coefficient corresponding a bin size of 4-14 minutes for a given channel. For Channel 1, this yielded $\beta$ values of 1.35, 1.38, and 1.65 for PMAP, PLD, and NNBR, respectively, and corresponding values for Channel 2 of 0.94, 1.08, and 1.15. While all methods performed extremely well by this metric, PMAP and PLD slightly outperformed NNBR.

We elected to use PMAP fits for all analyses, as it yielded both consistent transit depths in each channel and the lowest red noise levels. However, we highlight that all methods yielded broadly consistent transit parameters and there was no evidence for a systematic offset based on the fitting method used. For example, the difference in transit depth between NNBR and PMAP was 0.7$\sigma$ for 3.6\um, and 0.8$\sigma$ for 4.5\um\ observations, while the transit depth between PLD and PMAP was 1.6$\sigma$ for 3.6\um, and 2.7$\sigma$ for 4.5\um\ observations. The difference in the transit depth between PLD and PMAP in 4.5\um\ is much higher due to a single transit (transit number 193) from the PLD fit yielding an outlier result. If we removed this value, the transit depth between PLD and PMAP reduces to 1.8$\sigma$ for 4.5$\mu$m. Eccentricity results (Section~\ref{sec:ecc}) were similarly consistent across all methods used to correct the \spitzer\ light curves.

\begin{figure*}[htp]
    \includegraphics[width=0.48\linewidth]{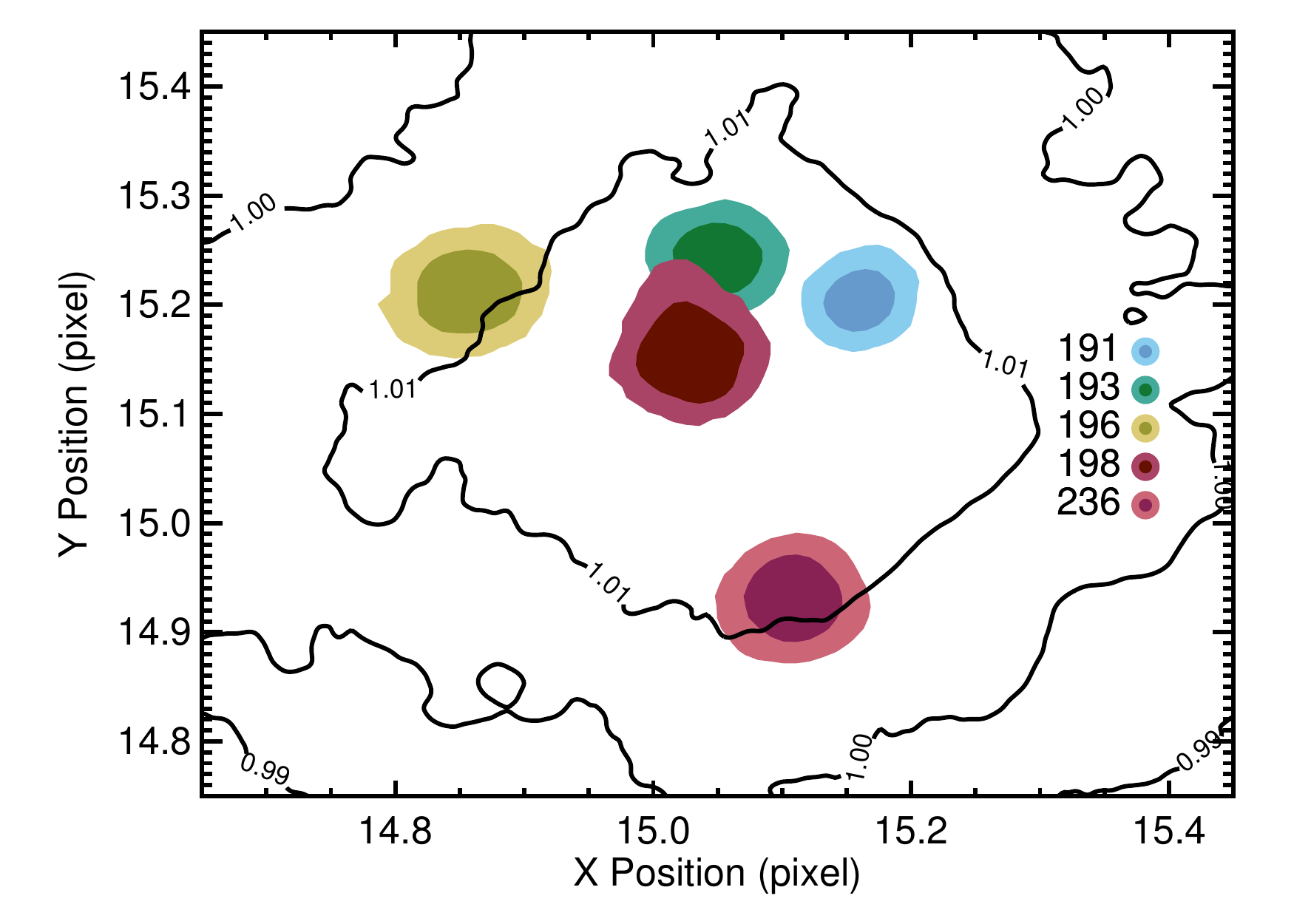}
    \includegraphics[width=0.5\linewidth]{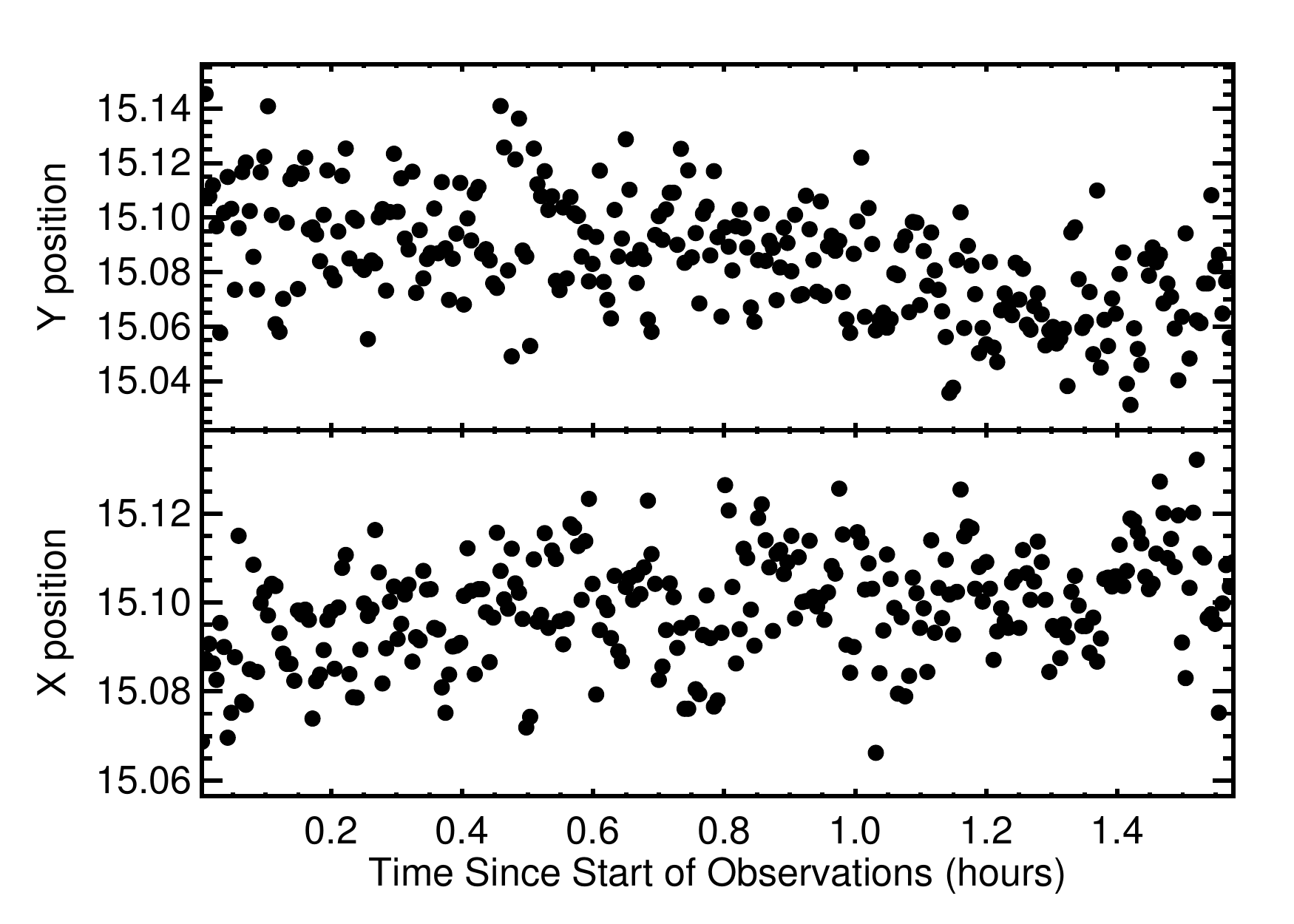}
    \caption{Left: distribution of $X$ and $Y$ pixel positions for each transit observed in Channel 2. Colored contours correspond to 68\% and 95\% of the centroid measurements, colored by the transit number (see Table~\ref{tab:obslog}). Black contours show the photometric response of IRAC Channel 2 as a function of centroid position from \citet{Ingalls2012}. Channel 1 is somewhat less well behaved -- only $>$75\% of the images had the PSF centroid within 0.25\,pixels of the sweet spot, while in Channel 2, this number was $>$90\%. Right: $X$ and $Y$ (pixel) centroid positions for transit 189 (Channel 1) over the observation. Centroid values are the average centroid of every 10 images for clarity and to mitigate random errors in centroiding. \label{fig:drift}}
\end{figure*}

Agreement between the  methods (except for the one transit in PLD), as well as the good performance of all the methods, were likely consequences of K2-2b's relatively short ($\simeq45$ minutes) transit duration and the small centroid drift ($\lesssim$0.1 pixels, Figure~\ref{fig:drift}a). Many similar studies utilized stares of 2-5 hr, which are subject to increased noise from long-term variability in \spitzer's temperature and CCD behavior. For observations of K2-25b, $>$75\% and $>$90\% of the images (for Channel 1 and 2, respectively) had the PSF centroid within 0.25\,pixels of the sweet spot. Pixel motion was also primarily random on timescales of the transit, adding more white than red noise to the light curve (Figure~\ref{fig:drift}b).

\subsection{MEarth}
We analyzed 12 transits of K2-25b using the \mearth-North and \mearth-South arrays \citep{Nutzman2008, Berta2013}. Details of these observations can be found in \cite{Kain2019}, which we briefly summarize here. \mearth-North used eight 40 cm telescopes at the Fred Lawrence Whipple Observatory on Mount Hopkins, Arizona. \mearth-South had a nearly identical set of telescopes located at the Cerro Tololo Inter-American Observatory (CTIO) in Chile. All telescopes used a $2048\times2048$ pixel CCD with pixels scales of 0.78\arcsec /pixel in the North and 0.84\arcsec /pixel in the South. Both telescope arrays used a Schott RG715 filter for all observations \citep[see][for the filter profile and CCD transmission]{2016ApJ...818..153D}. Observations analyzed here span 2015 December 9--2018 August 4. 

All telescopes integrated for 60\,s for a cadence of $\simeq$90\,s per telescope. In addition to transit observations, \mearth\ monitored K2-25 at regular intervals outside of the transit to better constrain the stellar variability. The observational strategy and analysis of this data are described in \cite{Newton2016} and \cite{Kain2019}.

MEarth data were reduced following the basic methodology from \citet{2007MNRAS.375.1449I} with additional steps detailed in the documentation of the fourth \mearth\ data release\footnote{\href{https://www.cfa.harvard.edu/MEarth/DR4/processing/index.html}{https://www.cfa.harvard.edu/MEarth/DR4/processing/index \ .html}}. This included corrections for second-order extinction (color differences between target and comparison stars), meridian flips (when the target crosses the meridian, the telescope rotates by 180$^{\circ}$ relative to the sky, and reference stars fall on different parts of the detector), and a fit to stellar variability derived from data taken out of transit. 

\subsection{LCOGT}

We observed two transits of K2-25b using the Las Cumbres Observatory Global Telescope network (LCOGT) 1 m telescope network \citep{2013PASP..125.1031B}. The first transit was observed on 2016 October 21 at the LCOGT node at McDonald Observatory in west Texas. The second transit was observed on 2016 October 28 simultaneously by two LCOGT telescopes at Cerro Tololo Inter-American Observatory (CTIO). Both observations used the Sinistro camera, SDSS $i$'-band filter, and with exposure times of 180 s. 

The LCOGT \textsc{banzai} pipeline \citep{McCully18} applied all basic data processing, including extraction for all sources into raw fluxes and uncertainties. To correct for atmospheric variability, we built a master comparison star from all stars detected in the LCOGT images for a given transit. We excluded sources within 20\% of saturation, K2-25b, and extended sources (identified by automated flags from the Banzai pipeline). We cross-matched each detection with AAVSO Photometric All-Sky Survey \citep[APASS;][]{APASS} and Sloan Digital Sky Survey \citep[SDSS;][]{Abolfathi2018} photometry and removed sources $>1.5$\,mag bluer than K2-25 in $V-r$ or $r-i$ color to minimize secondary extinction effects. We normalized the light curve from each comparison star (since we are interested in relative changes only), then combined them into a stacked curve using the robust weighted mean. We identified light curves exhibiting significant variability or trends inconsistent with the master curve by eye and repeated building the master curve from the remaining data. The flux measurements of K2-25 from \textsc{banzai} were then divided by this master curve. 

We removed an additional linear trend by fitting the out-of-transit data for each set of observations. The final linear trend may be due to stellar variability, shifts in the PSF over time, or airmass changes introducing weak color terms in the photometry \citep[e.g.,][]{Mann2011}.

\begin{deluxetable*} {lcccccc} 
\tabletypesize{\footnotesize} 
\tablecaption{Observation Log \label{tab:obslog}}
\tablecolumns{7}
\tablenum{1}
\tablewidth{0pt}
\tablehead{
\colhead{Date of First} & 
\colhead{Telescope/} &
\colhead{Number of } &
\colhead{Filter} & 
\colhead{Transit} & 
\colhead{Exposure} & 
\colhead{Number of} \\ 
\colhead{Exposure (UT)} & 
\colhead{Instrument} & 
\colhead{Telescopes } &
\colhead{} & 
\colhead{Number} & 
\colhead{Time (s)} & 
\colhead{Exposures}
}
\startdata
    2015 Feb 10 & \textit{K2} Campaign 4 & ... &  \textit{Kepler} & 1-20\tablenotemark{a} & 1765.5 & 3066 \\
    2015 Dec 9 & MEarth & 8 & RG715 & 87 & 60 & 2363 \\ 
    2015 Dec 16 & MEarth & 8 & RG715 & 89 & 60 & 2102 \\
    2016 Oct 7 & MEarth & 3& RG715 & 174\tablenotemark{b} & 60 & 547 \\ 
    2016 Oct 14 & MEarth & 5 & RG715 & 176 & 60 & 1348 \\ 
    2016 Oct 21 & MEarth & 5 & RG715 & 178 & 60 & 1309 \\ 
    2016 Oct 21 & LCOGT/Sinistro & ... &  SDSS $i$' & 178 & 180 & 101 \\ 
    2016 Oct 28 & MEarth & 4 & RG715 & 180 & 60 & 877 \\ 
    2016 Oct 28 & LCOGT/Sinistro & .. & SDSS $i$' & 180 & 180 & 177 \\
    2016 Nov 4 & MEarth & 4 & RG715 & 182 & 60 & 895 \\ 
    2016 Nov 11 & MEarth & 8 & RG715 & 184 & 60 & 1812 \\ 
    2016 Nov 18 & MEarth & 7 & RG715 & 186 & 60 & 1571 \\ 
    2016 Nov 25 & MEarth & 7 & RG715 & 188 & 60 & 1502 \\ 
    2016 Nov 28 & \textit{Spitzer}/IRAC & ... & Channel 1 & 189 & 2 & 2090 \\
    2016 Dec 2 & MEarth & 3 & RG715 & 190 & 60 & 915 \\ 
    2016 Dec 5 & \textit{Spitzer}/IRAC & ... & Channel 2 & 191 & 2 & 2544 \\
    2016 Dec 9 & \textit{Spitzer}/IRAC & ... & Channel 1 & 192 & 2 & 2547 \\
    2016 Dec 9 & MEarth & 3 & RG715 & 192 & 60 & 112 \\  
    2016 Dec 12 & \textit{Spitzer}/IRAC & ... & Channel 2 & 193 & 2 & 2409 \\
    2016 Dec 19 & \textit{Spitzer}/IRAC & ... &  Channel 1 & 195 & 2 &  2234 \\
    2016 Dec 22 & \textit{Spitzer}/IRAC &  ... & Channel 2 & 196 & 2 & 2429 \\
    2016 Dec 26 & \textit{Spitzer}/IRAC &  ... & Channel 1  & 197 &  2 & 2540 \\
    2016 Dec 29 &  \textit{Spitzer}/IRAC & ... &  Channel 2 & 198 & 2 & 2484 \\ 
    2017 Jan 2 & \textit{Spitzer}/IRAC &  ... & Channel 1 & 199 & 2 & 2542 \\
    2017 May 11 & \textit{Spitzer}/IRAC &  ... & Channel 2 & 236& 2 & 2208\\
\enddata
\tablenotetext{a}{ There were 20 total consecutive transits taken by \textit{K2}}
\tablenotetext{b}{ Only a partial transit was observed}
\end{deluxetable*}

\subsection{Timing corrections}

Our analysis included transit observations taken over $>2$ yr. While this provided extremely precise constraints on the period (subsecond precision), it makes it more critical that all observations were placed on the same time system. For easy comparison to the discovery data from \kepler, we converted all other data to Barycentric \textit{Kepler} Julian Day (BKJD), which is Barycentric dynamical time (BJD TBD) minus 2454833.0. For \spitzer, we followed the corrections given in the IRAC handbook\footnote{\href{https://irsa.ipac.caltech.edu/data/SPITZER/docs/irac/iracinstrumenthandbook/53/}{https://irsa.ipac.caltech.edu/data/SPITZER/docs/irac/irac \ instrumenthandbook/53/}} to place time in BJD TBD, with an additional small (1 s) correction to place the time at the center  of the integration, instead of the start of the integration. LCOGT time was given in UTC, which we convert to BJD following \citet{Eastman2010}\footnote{\href{http://astroutils.astronomy.ohio-state.edu/time/}{http://astroutils.astronomy.ohio-state.edu/time/}}. As with {\it Spitzer}, we applied a 90\,s correction to place the timestamp for LCOGT at the center of the integration. MEarth data already included a correction to the center of the exposure as detailed in the reduction documentation. 

\section{Updated Stellar Parameters from {\it Gaia}}\label{sec:params}

The availability of a precise (0.4\% error) parallax ($\pi$) from the second \textit{Gaia} data release \citep[DR2; ][]{GaiaDr2} enabled us to improve the stellar parameters presented in \cite{Mann2016a}, which relied on a less precise kinematic distance. To this end, we first computed $M_{K_S}$ from the inverted DR2 parallax and $K_S$ photometry from the Two Micron All Sky Survey \citep[2MASS; ][]{Skrutskie2006}. We then updated the stellar radius and mass values based on relations from \citet{Mann2015b} and \citet{Mann2019a}\footnote{\href{https://github.com/awmann/M_-M_K-}{\faGithub  https://github.com/awmann/M\_-M\_K-}}, adopting the metallicity of the Hyades \citep[e.g.,][]{Dutra-Ferreira2016, Liu2016}. Lastly, we combined the DR2 parallax with the bolometric flux determination from \citet{Mann2015b} to update the total stellar luminosity ($L_*$), and hence the effective temperature (\teff). 

The $M_*$ relation from \citet{Mann2019a} is empirically calibrated using dynamical mass measurements of astrometric binaries and resolved $K_S$ magnitudes from adaptive optics imaging. The $R_*$ relation from \citet{Mann2015b} used stellar atmosphere models to compute \teff, which was then converted to $R_*$ using the Stefan-Boltzmann relation and measurements of the total luminosity from absolutely calibrated spectra and literature distances. However, the model-based \teff\ determinations were calibrated (down-weighting spectral regions poorly reproduced by models) using empirical determinations from long-baseline optical interferometry \citep{Boyajian2012,Mann2013c}. Updated empirical relations based on a larger grid of interferometric radii yielded consistent results \citep{2019MNRAS.484.2674R}.

Uncertainties in $M_*$ and $R_*$ account for both measurement errors (in $K_S$ and parallax) and uncertainties in the calibrations. Our final adopted parameters are listed in Table~\ref{tab:compare} (middle column). The updated parameters are consistent with the original determination, but more precise than those from the discovery paper (right column of Table~\ref{tab:compare}). 

\begin{deluxetable} {l|rr} 
\tablecaption{Updated Stellar Parameters
\label{tab:compare}}
\tablecolumns{3}
\tablenum{2}
\tablewidth{0pt}
\tablehead{
\colhead{Parameters} & 
\colhead{This Work } & 
\colhead{\cite{Mann2016a}}
}
\startdata
    $R_*$ $(R_\odot)$ & $0.2932\pm0.0093$ & $0.295\pm0.020$\\
    $M_*$ $(M_\odot)$ & $0.2634\pm0.0077$ &  $0.294\pm0.021$\\
    $\rho_*$ $(\rho_\odot)$ & $10.45\pm0.73$ & $11.3\pm1.6$\\
    $L_*$ $(L_\odot) $ & $8.16\pm0.29\times10^{-3}$ & $8.4\pm1.4\times10^{-3}$ \\
    $\teff$ (K) & $3207\pm58$ & $3180\pm60$ \\
\enddata
\end{deluxetable}

While the relations we utilized to derive updated parameters for K2-25 were built from older M dwarfs ($\gtrsim1$\,Gyr), mid-M dwarfs like K2-25 arrive at the main sequence around $\simeq 100$\,Myr \citep[e.g.,][]{baraffe2015new}, much younger than the age of Hyades. We discuss the effects of activity on our results further in Section~\ref{sec:ecc}.

\section{Transit Fitting}\label{sec:transit}
Our transit-fitting procedure followed the same steps from \citet{Mann2016b}, with the exception that we fit each dataset (wavelength/instrument) separately. We briefly summarize our transit fitting method below.

We fit the extracted light curves to transit models within an MCMC, using a modified version of the \texttt{misttborn} code\footnote{\url{https://github.com/captain-exoplanet/misttborn}}. For this, we utilized the \textit{emcee} Python module \citep{Foreman-Mackey2013} and the \textit{batman} package to generate the transit models, which uses the \citet{MandelAgol2002} transit model. The nine free parameters explored in the MCMC were the planet-to-star radius ratio ($R_{p}/R_{*}$), impact parameter (\textit{b}), orbital period (\textit{P}), epoch of the first transit midpoint (\textit{$T_{0}$}), two parameters that describe the eccentricity and argument of periastron (\textit{$\sqrt{e}$ sin($\omega$)} and \textit{$\sqrt{e}$ cos($\omega$})), bulk stellar density ($\rho_
{*}$), and two limb-darkening parameters (see below). We assumed a linear ephemeris, but a parallel search for transit timing variations showed that this a reasonable assumption \citep{Kain2019}. For the PLD fits to the \spitzer\ data, we included additional free parameters to describe the intra-pixel sensitivity variations. Each MCMC chain was run using 100 walkers for 200,000 steps. 

For limb-darkening, we used the triangular limb-darkening parameters ($q_1$ and $q_2$) described by \citet{Kipping2013} to uniformly explore the physically allowed region of parameter space. For {\it K2}, MEarth, and LCOGT data, we placed Gaussian priors on limb-darkening parameters derived using LDTK \citep{2015MNRAS.453.3821P}, which estimates limb-darkening from the \citet{2013A&A...553A...6H} stellar atmosphere models. For the {\it Spitzer} bands, we used limb-darkening parameters estimated by \citet{Claret2011}. We adopted uncertainties on these limb-darkening parameters based on both errors in the stellar parameters (see Section~\ref{sec:params}), and differences in values based on the model-grid and interpolation method used. This resulted in Gaussian prior widths of 0.03-0.05, depending on the wavelength. The linear and quadratic limb-darkening coefficient priors ($g_{1}$ and $g_{2}$), along with the wavelength range, mean wavelength ($\lambda_{\rm{mean}}$), and effective wavelength ($\lambda_{\rm{eff}}$) for each data set are presented in Table~\ref{tab:limb_darkening_priors}. Our fit used the triangular limb-darkening coefficients, but linear and quadratic darkening coefficients are listed in Table~\ref{tab:limb_darkening_priors} for reference.

\begin{deluxetable} {llclrr}
\tabletypesize{\scriptsize}
\tablecaption{Priors on Limb-darkening Coefficients \label{tab:limb_darkening_priors}}
\tablecolumns{6}
\tablenum{3}
\tablewidth{0pt}
\tablehead{
\colhead{Telescope} & 
\colhead{$\lambda$ Range} &
\colhead{$\lambda_{\rm{mean}}$} &
\colhead{$\lambda_{\rm{eff}}$} &
\colhead{$g_{1}$} &  
\colhead{$g_{2}$}  \\ 
\colhead{} & 
\colhead{($\mu$m)} & 
\colhead{($\mu$m)} & 
\colhead{($\mu$m)} & 
\colhead{} & 
\colhead{} 
}
\startdata 
    \textit{K2} & 0.42--0.90 & 0.64 & 0.73 & $0.42\pm0.03$ & $0.31\pm0.04$ \\
    LCOGT& 0.66--0.85 & 0.76 & 0.77 &  $0.34\pm0.05$ & $0.34\pm0.05$ \\ 
    MEarth & 0.69--1.00 & 0.83 & 0.84 & $0.28\pm0.05$ & $0.33\pm0.05$\\ 
    Channel 1 & 3.13--3.96 & 3.56 & 3.46 & $0.06\pm0.03$ & $0.19\pm0.04$\\
    Channel 2 & 3.92--5.06 & 4.50 & 4.43 & $0.06\pm0.03$ & $0.16\pm0.04$ \\
\enddata
\tablecomments{\textit{K2}, LCOGT, and MEarth values were calculated using the LDTK toolkit \citep{2015MNRAS.453.3821P} and the \textit{Spitzer} bands were calculated using that of \citet{Claret2011}. Limb-darkening priors are provided as the traditional linear and quadratic terms, but were fit using triangular sampling terms.}
\end{deluxetable}

The lower cadence (30 minutes) provided by the \ktwo\ and the poorer precision in the LCOGT photometry meant that these two datasets provided only limited constraints on the impact parameter. To solve this, we added a Gaussian prior on the impact parameter for these two datasets derived from the \textit{Spitzer} Channel 2 fit ($b=0.658\pm0.043$). For all other datasets, we used uniform priors on impact parameter, as their cadence was sufficient for similarly precise constraints on impact parameter. We applied a Gaussian prior on the stellar density taken from our analysis in Section~\ref{sec:params} for all light curves. All other parameters ($R_{p}/R_{*}$, $P$, $T_{0}$, \textit{$\sqrt{e}$ sin($\omega$)}, and \textit{$\sqrt{e}$ cos($\omega$})) were fit using uniform priors within physically allowed bounds (e.g., $P$>0).

The autocorrelation time was $<$2500 steps for all fits, with an effective number of samples across all walkers of $>$8000 (80 per walker), which was more than sufficient for convergence in all cases.

Results of each fit are provided in Table~\ref{tab:param} and the model light curves with the best-fit parameters (highest likelihood) for each dataset is shown in Figure~\ref{fig:lc_all}. 

To better constrain the wavelength-independent parameters of K2-25b, we also ran an MCMC fit that combined all the datasets together (``combined fit''). This fit comprised of 17 total free parameters, with 10 of those being limb-darkening parameters (two for each wavelength). We applied the same priors on limb-darkening and stellar density used in the individual fits for each wavelength. Unlike the individual \textit{K2} and LCOGT fits, we did not apply a Gaussian prior on the impact parameter in these two datasets. This combined fit was particularly useful for constraining the impact parameter and the orbital period of the planet. The results from the MCMC fit using all of the data are presented in Table~\ref{tab:gfparam}, Fit 1.

\section{Results}\label{sec:results} 

\subsection{Transit parameters}

\begin{figure*} [!ht]
    \centering
    \includegraphics[width=.75\textwidth,height=.75\textwidth]{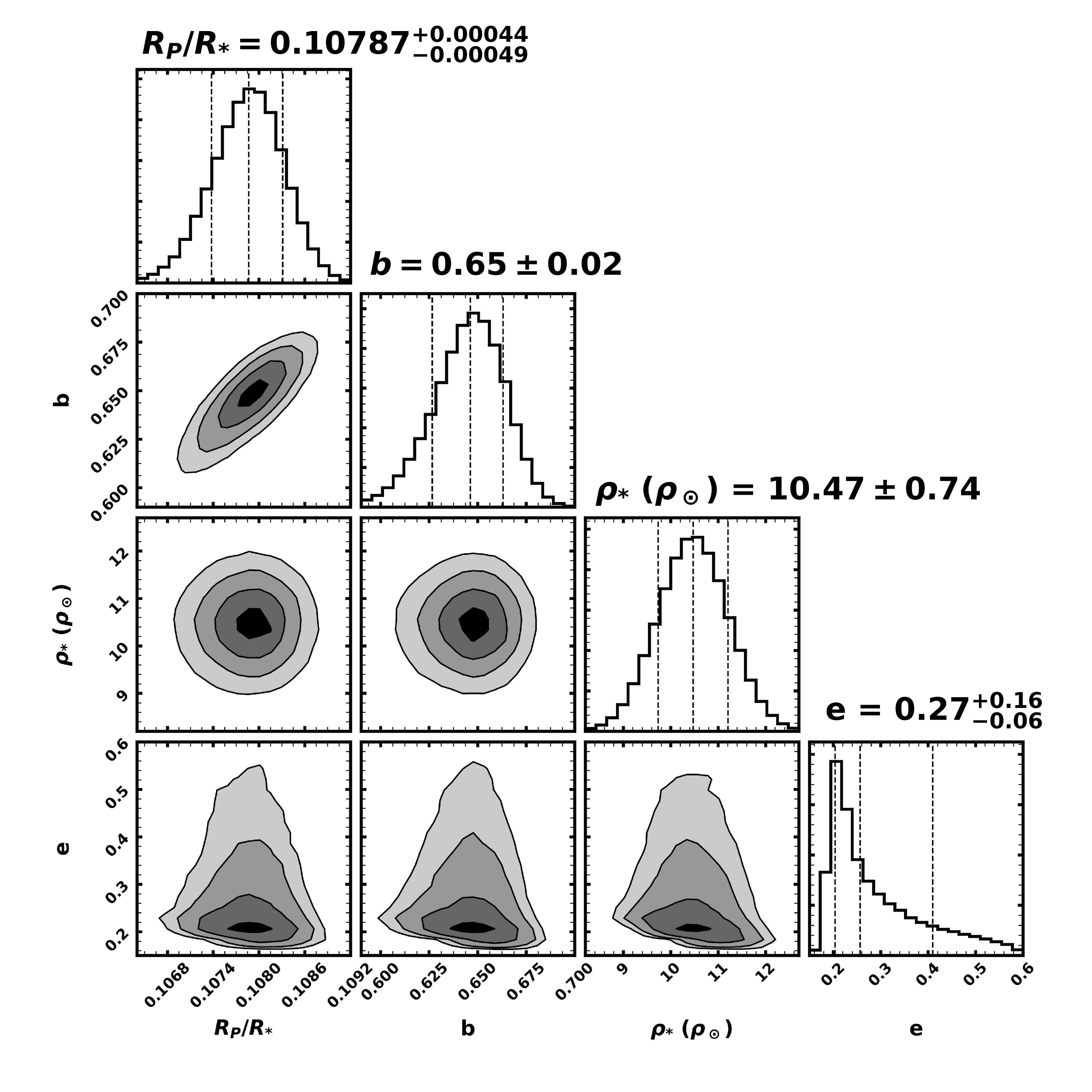}
    \caption{Posteriors from the MCMC fit using all of the datasets for the planet-to-star radius ratio ($R_{p}/R_{*}$), impact parameter ($b$), stellar density ($\rho_{*}$), and eccentricity ($e$). In each histogram, the dashed lines indicate the 16\%, 50\%, and 84\% percentiles. Plot was created by \texttt{corner.py} \citep{foreman2016corner}\label{fig:corner}.}  
\end{figure*}

The measured parameters, as well as the derived parameters (semi-major axis (\textit{a}), eccentricity ($e$), radius of planet ($R_{p}$), argument of periastron ($\omega$), triangular limb-darkening parameters (\textit{$q_{1}$} and \textit{$q_{2}$}), semi-major axis ratio ($a/R_{*}$), transit depth ($\delta$), and orbital inclination ($i$)), from our MCMC fit for each dataset or wavelength are given in Table~\ref{tab:param}. The final parameters from the combined fit (combined datasets) are given in Fit 1 of  Table~\ref{tab:gfparam}, with the posteriors and correlations shown in Figure~\ref{fig:corner}.

\begin{deluxetable*} {l|lllll}
\tabletypesize{\footnotesize}
\tablecaption{Transit Fit Parameters \label{tab:param}}
\tablecolumns{6}
\tablenum{4}
\tablewidth{0pt}
\tablehead{
\colhead{Parameter} & 
\colhead{\textit{K2}} &
\colhead{MEarth} & 
\colhead{LCOGT} & 
\colhead {Channel 1} & 
\colhead {Channel 2}  
}
\startdata 
\multicolumn{6}{c}{\emph{Measured Parameters}} \\
\hline
\decimals
\rotate
    Orbital period, $P$ (days) & $3.484545^{+4.2\times10^{-5}}_{-4.3\times10^{-5}}$ & $3.4845617 \pm 1.7\times10^{-6}$ & $3.48423 \pm 0.00023$ & $3.484552^{+1.9\times10^{-5}}_{-1.8\times10^{-5}}$ & $3.4845645 \pm 4.7\times10^{-6}$\\
    Planet-to-star radius ratio, $R_{P}/R_{*}$ & $0.1095^{+0.0025}_{-0.0023}$ & $0.10912^{+0.00082}_{-0.00097}$ & $0.1099^{+0.0021}_{-0.0019}$ &  $0.1069^{+0.00070}_{-0.00075}$ & $0.10759^{+0.00077}_{-0.00084}$ \\
    Epoch of first transit midpoint, $T_{0}$ (BJD-2400000)\tablenotemark{a} & $57062.5795^{+0.00054}_{-0.00056}$  & $57062.5799^{+0.00031}_{-0.00032}$ & $57062.5799^{+0.00031}_{-0.00032}$ & $57062.5816^{+0.0035}_{-0.0036}$ & 
    $57062.5794\pm 0.00097$ \\
    Impact parameter, \textit{b} & $0.667^{+0.044}_{-0.043}$ & $0.671^{+0.024}_{-0.032}$ &  $0.664^{+0.039}_{-0.040}$ & 
    $0.643^{+0.032}_{-0.036}$ & $0.658^{+0.033}_{-0.043}$ \\
    Stellar density, $\rho_{*}$ $(\rho_\odot)$ & $10.482^{+ 0.728}_{-0.721}$ & $10.468^{+0.719}_{-0.728}$ & $10.479^{+ 0.723}_{-0.726}$ & $10.481^{+0.726}_{-0.730}$ & $ 10.478^{+ 0.727}_{-0.724} $\\
    Triangular limb-darkening coefficient, $q_{1}$ & $0.545^{+0.070}_{-0.071}$ & $0.303^{+0.079}_{-0.071}$ &  $0.505^{+0.093}_{-0.092}$ & $0.092^{+0.050}_{-0.044}$ &  $0.086^{+0.051}_{-0.043}$ \\ 
    Triangular limb-darkening coefficient, $q_{2}$ & $0.288 \pm 0.028$ & $0.227^{+0.047}_{-0.048}$ & $0.26 \pm 0.045$ & $0.247^{+0.070}_{-0.065}$ & $0.142^{+0.075}_{-0.070}$ \\
    \textit{$\sqrt{e}$ sin($\omega$)} & $0.26^{+0.15}_{-0.34}$ & $0.322^{+0.09}_{-0.20}$ & $0.31^{+0.11}_{-0.23}$ & $0.358^{+0.08}_{-0.20}$ & $0.339^{+0.089}_{-0.200}$ \\ 
    \textit{$\sqrt{e}$ cos($\omega$)} & $-0.01 \pm 0.65$ & $-0.13^{+0.53}_{-0.50}$ & $-0.08^{+0.58}_{-0.56}$ & $-0.15 \pm 0.49$ & $-0.18^{+0.52}_{-0.49}$ \\
    \hline 
    \multicolumn{6}{c}{\emph{Derived Parameters}}\\
    \hline
    Eccentricity, \textit{e} & $0.32^{+0.31}_{-0.11}$ &  $0.255^{+0.166}_{-0.065}$ & $0.285^{+0.209}_{-0.081}$ & $0.263^{+0.170}_{-0.057}$ & $0.265^{+0.191}_{-0.064}$ \\
    Planet radius, $R_{p}$ ($R_{\oplus})$ & $3.497 \pm 0.135$ & $3.486 \pm 0.112$ & $3.520^{+0.135}_{-0.123}$ & $3.418 \pm 0.111$ & $3.440\pm0.112$ \\ 
    Argument of periastron, $\omega$ ($^{\circ}$) & $118.0^{+74.0}_{-87.0}$ & $107.0^{+60.0}_{-69.0}$ & $103.0^{+66.0}_{-73.0}$ & $109.0^{+57.0}_{-62.0}$ &  $113.0^{+56.0}_{-67.0}$ \\
    Regular limb-darkening coefficient, $g_{1}$ & $0.424^{+0.052}_{-0.049}$ & $0.247^{+0.058}_{-0.055}$ & $0.366^{+0.074}_{-0.070}$ & $0.092^{+0.050}_{-0.044}$ & $0.078^{+0.053}_{-0.042}$ \\ 
    Regular limb-darkening coefficient, $g_{2}$ & $0.311^{+0.047}_{-0.045}$ & $0.297^{+0.073}_{-0.065}$ & $0.338^{+0.073}_{-0.069}$  & $0.247^{+0.070}_{-0.065}$ & $0.202^{+0.077}_{-0.069}$ \\ 
    Semi-major Axis Ratio, $a/R_{*}$ & $25.63^{+0.83}_{-0.86}$ & $25.41^{+0.51}_{-0.48}$ & $25.73^{+0.66}_{-0.71}$ & $25.93 \pm 0.56$ & $25.75^{+0.61}_{-0.58}$\\
    Transit depth, $\delta$ (\%) & $1.200^{+0.056}_{-0.049}$ & $1.191^{+0.018}_{-0.021}$ & $1.209^{+0.046}_{-0.042}$ & $1.143^{+0.015}_{-0.016}$ & $1.158^{+0.017}_{-0.018}$ \\ 
    Inclination, $i$ ($^{\circ}$) & $88.05^{+0.17}_{-0.40}$ & $88.1^{+0.10}_{-0.14}$ & $88.09^{+0.14}_{-0.20}$ & $88.17^{+0.12}_{-0.15}$ & $88.13^{+0.13}_{-0.17}$\\
\enddata
\tablenotetext{a}{ BJD is given in Barycentric Dynamical Time (TBD) format}
\end{deluxetable*}

\clearpage

\begin{figure*}[htp!]
    \includegraphics[width=0.975\textwidth]{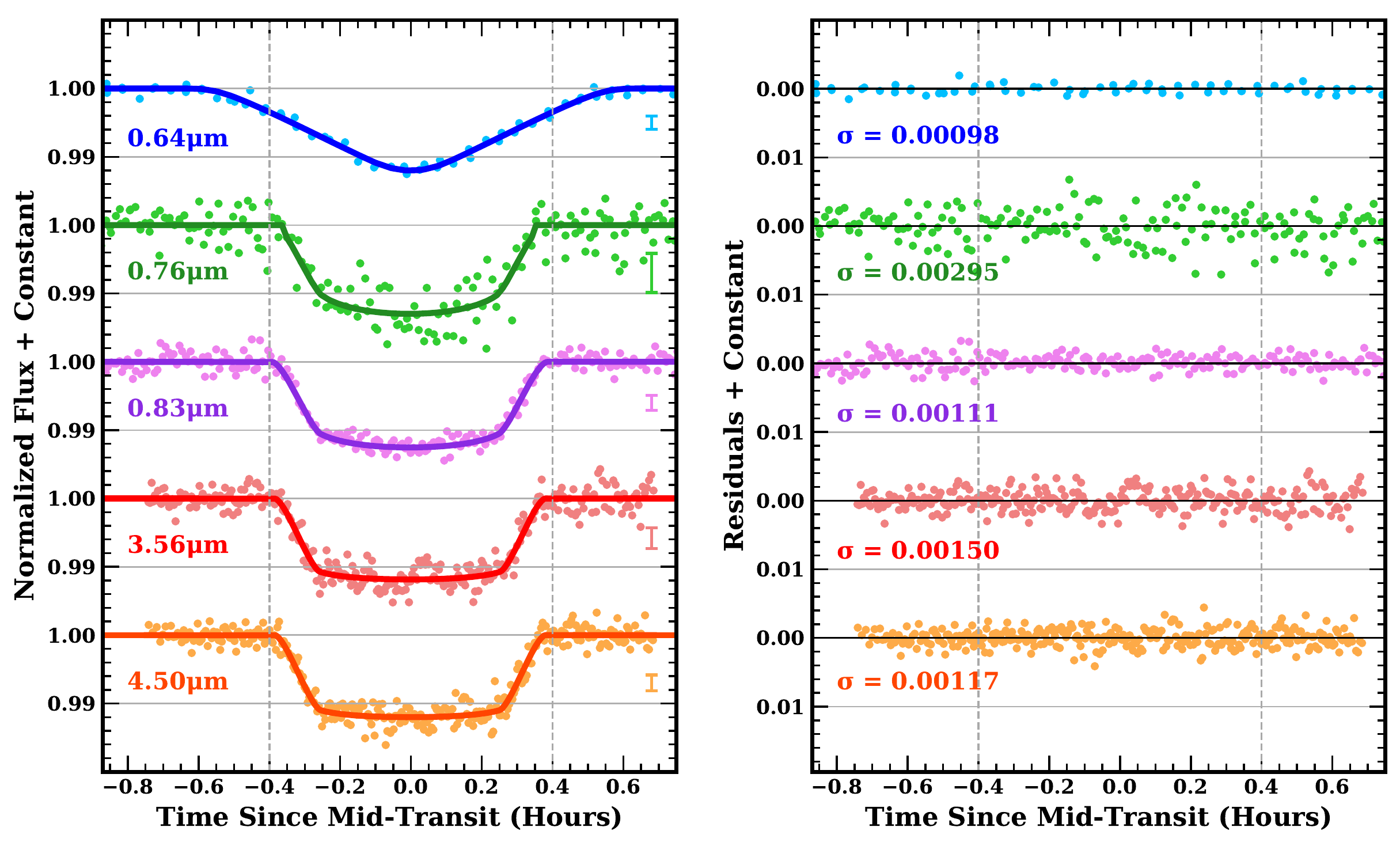}
    \caption{Left: phase-folded light curves of K2-25's transit observed in the mean filter wavelengths of 0.64$\mu$m (blue: \kepler; 20 transits), 0.76$\mu$m (green: LCOGT; 2 transits), 0.83$\mu$m (purple: MEarth; 12 transits), 3.56$\mu$m (red: \spitzer; 5 transits), and 4.50$\mu$m (orange: \spitzer; 5 transits). The \spitzer\ and MEarth data are binned using a median bin size of 80 s and $\sim$ 360 s, respectively. The solid line is the best-fit model, convolved with the integration time of each observation. The dashed gray vertical line at x = $\pm$ 0.4 hr is plotted for visual aid. Typical error bars are derived from scatter in the out-of-transit data displayed points and are displayed at the right side of the light curves. Right: residuals from the best fits with RMS of the displayed points}. \label{fig:lc_all}
\end{figure*}

\begin{deluxetable} {l|ll} 
\tabletypesize{\footnotesize}
\tablecaption{Combined Fit \label{tab:gfparam}}
\tablecolumns{3}
\tablenum{5}
\tablewidth{0pt}
\tablehead{
\colhead{Parameter} & 
\colhead{Fit 1\tablenotemark{a} (Preferred)} &
\colhead{Fit 2\tablenotemark{a} ($e$ = 0)} 
}
\startdata 
\multicolumn{3}{c}{\emph{Measured Parameters}} \\ 
\hline
\decimals
    $P$ (days) & $3.48456322^{+9.7\times10^{-7}}_{-9.5\times10^{-7}}$ & $3.48456325^{+9.7\times10^{-7}}_{-9.6\times10^{-7}}$ \\
    $R_{P}/R_{*}$ & $0.10787^{+0.00044}_{-0.00049}$ & $0.10784^{+0.00045}_{-0.00050}$ \\
    $T_{0}$ (BJD-2400000)\tablenotemark{b} & $57062.57965 \pm 0.0002$ & $57062.57965\pm 0.00018$ \\
    \textit{b} & $0.646^{+0.017}_{-0.020}$ & $0.645^{+0.017}_{-0.020}$ \\
    $\rho_{*}$ $(\rho_\odot)$ & $10.471^{+0.739}_{-0.735}$ & $ 34.921^{+1.925}_{-2.165}$\\
    $q_{1}$ \ktwo\ & $0.549^{+0.094}_{-0.093}$ & $0.548^{+0.095}_{-0.094}$ \\ 
    $q_{2}$ \ktwo\ & $0.285 \pm 0.043$ & $0.285 \pm 0.043$ \\ 
    $q_{1}$ MEarth & $0.282^{+0.056}_{-0.053}$ & $0.282^{+0.056}_{-0.053}$ \\ 
    $q_{2}$ MEarth & $0.201 \pm 0.047$  & $0.202 \pm 0.047$ \\ 
    $q_{1}$ LCOGT & $0.518 \pm 0.084$ & $0.517^{+0.085}_{-0.084}$ \\ 
    $q_{2}$ LCOGT & $0.254^{+0.044}_{-0.045}$ & $0.256 \pm 0.044$ \\ 
    $q_{1}$ \spitzer\ Channel 1 & $0.135^{+0.040}_{-0.036}$ & $0.135^{+0.040}_{-0.036}$  \\ 
    $q_{2}$ \spitzer\ Channel 1  & $0.165 \pm 0.063$ & $0.166 \pm 0.063$ \\
    $q_{1}$ \spitzer\ Channel 2 & $0.105^{+0.040}_{-0.036}$ &  $0.106^{+0.040}_{-0.035}$ \\ 
    $q_{2}$ \spitzer\ Channel 2 & $0.156^{+0.073}_{-0.072}$ &  $0.157^{+0.074}_{-0.072}$ \\ 
    \textit{$\sqrt{e}$ sin($\omega$)} & $0.341^{+0.091}_{-0.200}$ &  0 (fixed) \\ 
    \textit{$\sqrt{e}$ cos($\omega$)} & $-0.05^{+0.55}_{-0.49}$ & 0 (fixed) \\
    \hline 
    \multicolumn{3}{c}{\emph{Derived Parameters}}\\
    \hline
    \textit{e} & $0.27^{+0.163}_{-0.064}$ & 0 (fixed) \\
    $R_{p}$ ($R_{\oplus})$ & $3.4492^{+0.1099}_{-0.1110}$ & $3.4480^{+0.1099}_{-0.1110}$ \\ 
    $\omega$ ($^{\circ}$) & $98.0^{+59.0}_{-68.0}$ & 0 (fixed) \\
    $g_{1}$ \ktwo\ & $0.419^{+0.072}_{-0.069}$ &  $0.419^{+0.074}_{-0.070}$ \\ 
    $g_{2}$ \ktwo\ & $0.316^{+0.072}_{-0.069}$ & $0.315^{+0.072}_{-0.068}$ \\ 
    $g_{1}$ MEarth & $0.211^{+0.051}_{-0.049}$ & $0.213^{+0.051}_{-0.049}$ \\ 
    $g_{2}$ MEarth & $0.315^{+0.065}_{-0.061}$ & $0.315^{+0.066}_{-0.061}$ \\ 
    $g_{1}$ LCOGT & $0.363^{+0.071}_{-0.068}$ & $0.365^{+0.070}_{-0.068}$ \\
    $g_{2}$ LCOGT & $0.352^{+0.072}_{-0.070}$ & $0.348^{+0.073}_{-0.069}$ \\
    $g_{1}$ \spitzer\ Channel 1 & $0.119^{+0.051}_{-0.047}$ & $0.12^{+0.050}_{-0.047}$ \\ 
    $g_{2}$ \spitzer\ Channel 1 & $0.242^{+0.062}_{-0.055}$ & $0.241^{+0.062}_{-0.056}$ \\ 
    $g_{1}$ \spitzer\ Channel 2 & $0.098^{+0.053}_{-0.046}$ & $0.099^{+0.053}_{-0.047}$  \\ 
    $g_{2}$ \spitzer\ Channel 2 & $0.217^{+0.066}_{-0.059}$ & $0.218^{+0.067}_{-0.060}$ \\ 
    $a/R_{\star}$ & $25.85 \pm 0.39$ & $31.61^{+0.64}_{-0.59}$ \\
    $\delta$ (\%) &    $1.1635^{+0.0096}_{-0.0100}$ & $1.163^{+0.0098}_{-0.0100}$\\
    $i$ ($^{\circ}$) & $88.164^{+0.085}_{-0.100}$ & $88.831^{+0.059}_{-0.053}$ \\ 
\enddata
\tablenotetext{a}{All fits were done with priors on $q_{1}$ and $q_{2}$ (different values for each dataset or wavelength), Fit 1 included a prior on $\rho_{*}$, and Fit 2 was done with \textit{e} and \textit{$\omega$} fixed at 0 and a uniform prior on $\rho_{*}$.} 
\tablenotetext{b}{ BJD is given in Barycentric Dynamical Time (TBD) format}
\end{deluxetable}

Across all data sources, the wavelength-independent parameters (e.g., $P$, $T_0$) are consistent. Results from the \spitzer\ bands had the tightest constraints due to the combination of high cadence (2 s) and precise photometry. 

 We were able to significantly improve the ephemeris of K2-25b, providing subminute level transit predictions well into the next decade. We also measured the planet radius to 3\% ($3.45\pm 0.11R_{\oplus}$), which is sufficient to characterize the density (and hence composition) of the planet when a comparably precise ($\simeq$10\%) mass determination becomes available. Our fits also supports a nonzero eccentricity ($e=0.27^{+0.16}_{-0.06}$), suggested by the unusually short transit duration.

\subsubsection{Eccentricity}\label{sec:ecc}

Our transit fit posteriors indicated a large eccentricity ($>0.2$) for K2-25b's orbit when compared to earlier studies suggesting close-in planets orbiting older stars have small orbital eccentricities \citep[0-0.15; e.g.,][]{2013ApJ...772...74W, Van-Eylen2015, Hadden2017, Mann2017b}. 

\begin{figure}[ht!]
    \hfill\includegraphics[width=0.47\textwidth]{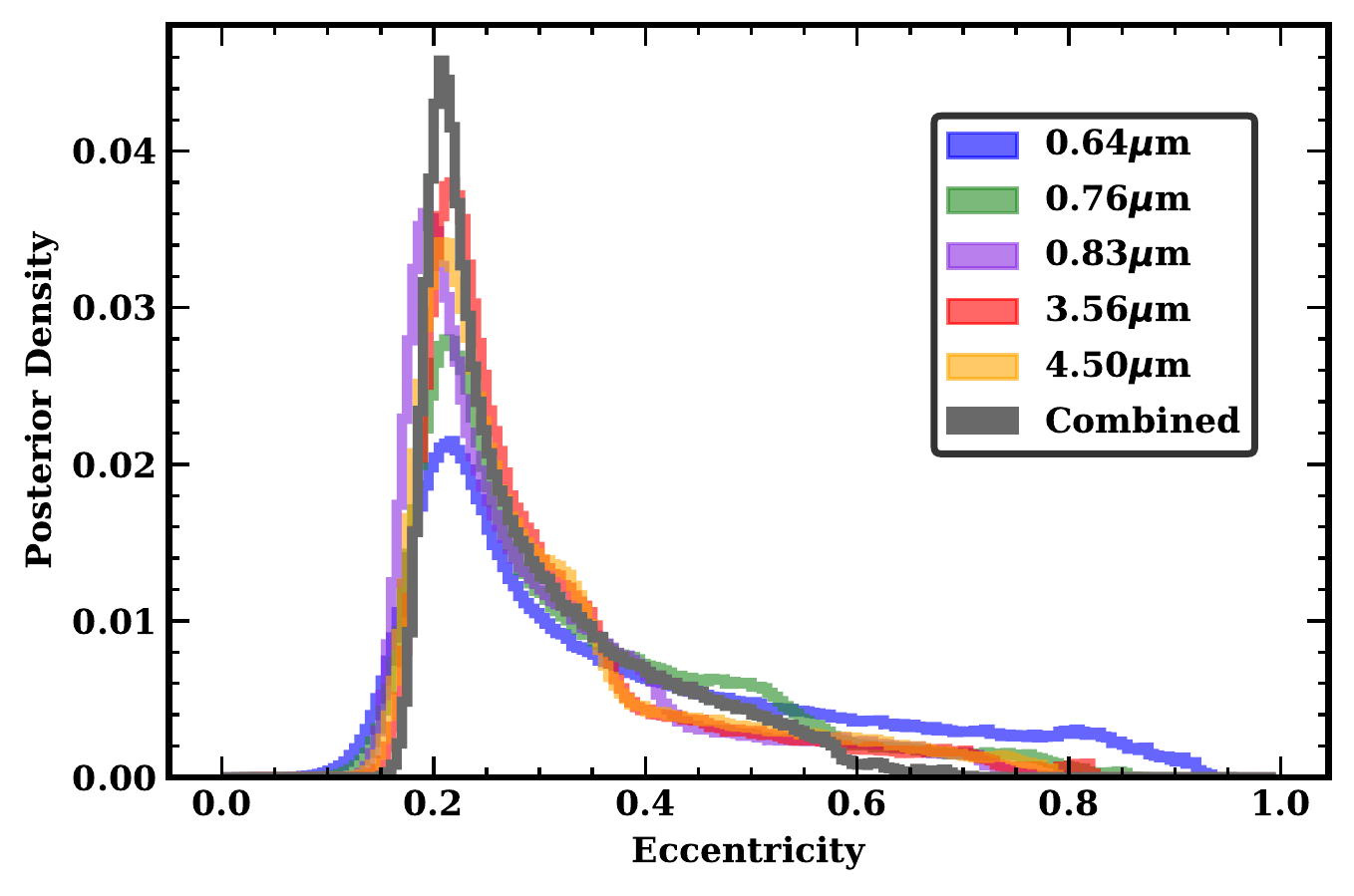}
    \caption{Eccentricity posteriors from the MCMC fits colored by data source (and wavelength) with a bin width of 0.005. The consistency between these fits rules out stellar activity or systematic errors in the photometry as potential sources of the large eccentricity. \label{fig:epost}}
\end{figure}

The large eccentricity result was independent of the dataset ($K2$, LCOGT, MEarth, and \spitzer) used (Figure~\ref{fig:epost} and Table~\ref{tab:param}), indicating the result was robust to stellar signals and systematics in the transit photometry. The spot contrast and flare strength decrease with increasing wavelength; if either were biasing the measured transit duration (and hence eccentricity determination), we would have expected to have a smaller eccentricity value from the longest-wavelength data. Similarly, each source of photometry was subject to different sources of systematic/correlated noise (e.g., intra-pixel variations for \spitzer\ versus atmospheric transparency variations from MEarth). If poor correction of these effects were behind the eccentricity results, we would see significant differences as a function of the instrument.

\begin{figure}[ht!]
    \hfill\includegraphics[width=0.47\textwidth]{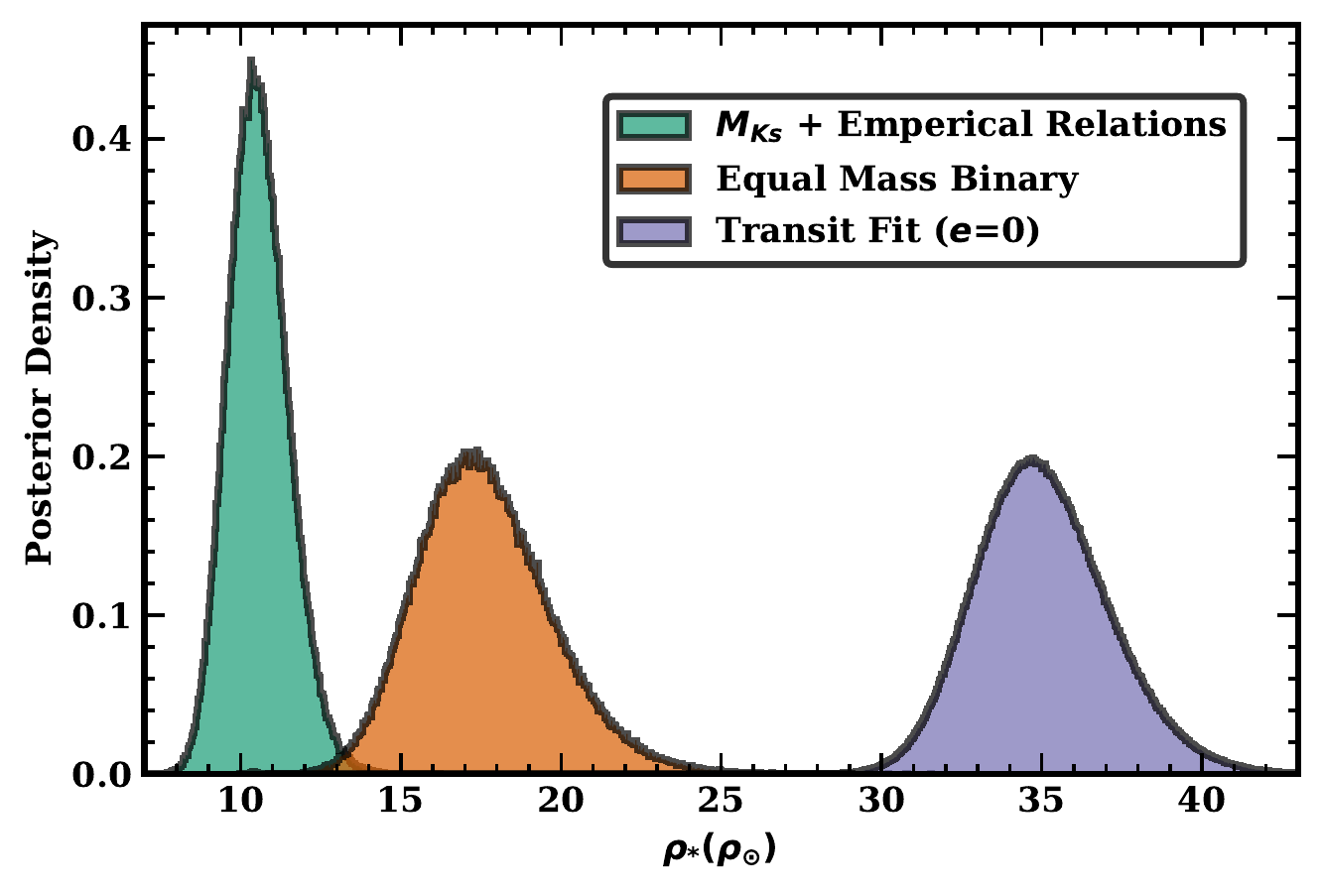}
    \caption{Comparison of the stellar density ($\rho_{*}$) from the transit fit posterior assuming $e=0$ (purple) to the density posterior derived from the \textit{Gaia} distance, $M_{K_S}-M_*$, and $M_{K_S}-R_*$ relations (green; Section~\ref{sec:params}) and the density posterior using the same relations but assuming the star is a unresolved equal-mass binary (orange). \label{fig:dencomp}}
\end{figure}

One feature common to all fits that could impact the derived eccentricity was the Gaussian stellar density prior. As an additional test of this, we reran the combined MCMC transit fit as described in Section~\ref{sec:transit}, but with eccentricity fixed at zero and a uniform prior on stellar density. The resulting fit yielded a stellar density of $35\pm2\rho_{\odot}$, which is inconsistent with the independent stellar density derived from the DR2 distance and empirically calibrated mass-luminosity and mass-radius relations (see Section~\ref{sec:params}) at $>10\sigma$, as we show in Figure~\ref{fig:dencomp}. The results of this fit are presented in Table~\ref{tab:gfparam}, Fit 2. 

The relations used to assign a stellar density for K2-25 were based on stars that are generally older and less active than K2-25. Some studies have suggested that younger and/or more active M dwarfs are larger than their older counterparts \citep[e.g.,][]{Stassun2012, 2017ApJ...845...72K, Jaehnig2018}, and some are not \citep[e.g.,][]{Kesseli2018, Jackson2019}. At most, this would increase the inferred mass (relative to the true mass) by $<5\%$, and the radius by $<10\%$. The net effect would make K2-25 {\it less dense} by $\simeq20\%$, increasing the discrepancy between the transit-fit and luminosity-based densities or requiring an even larger eccentricity for the orbit. 

\begin{figure}[t]
    \hfill\includegraphics[width=0.47\textwidth]{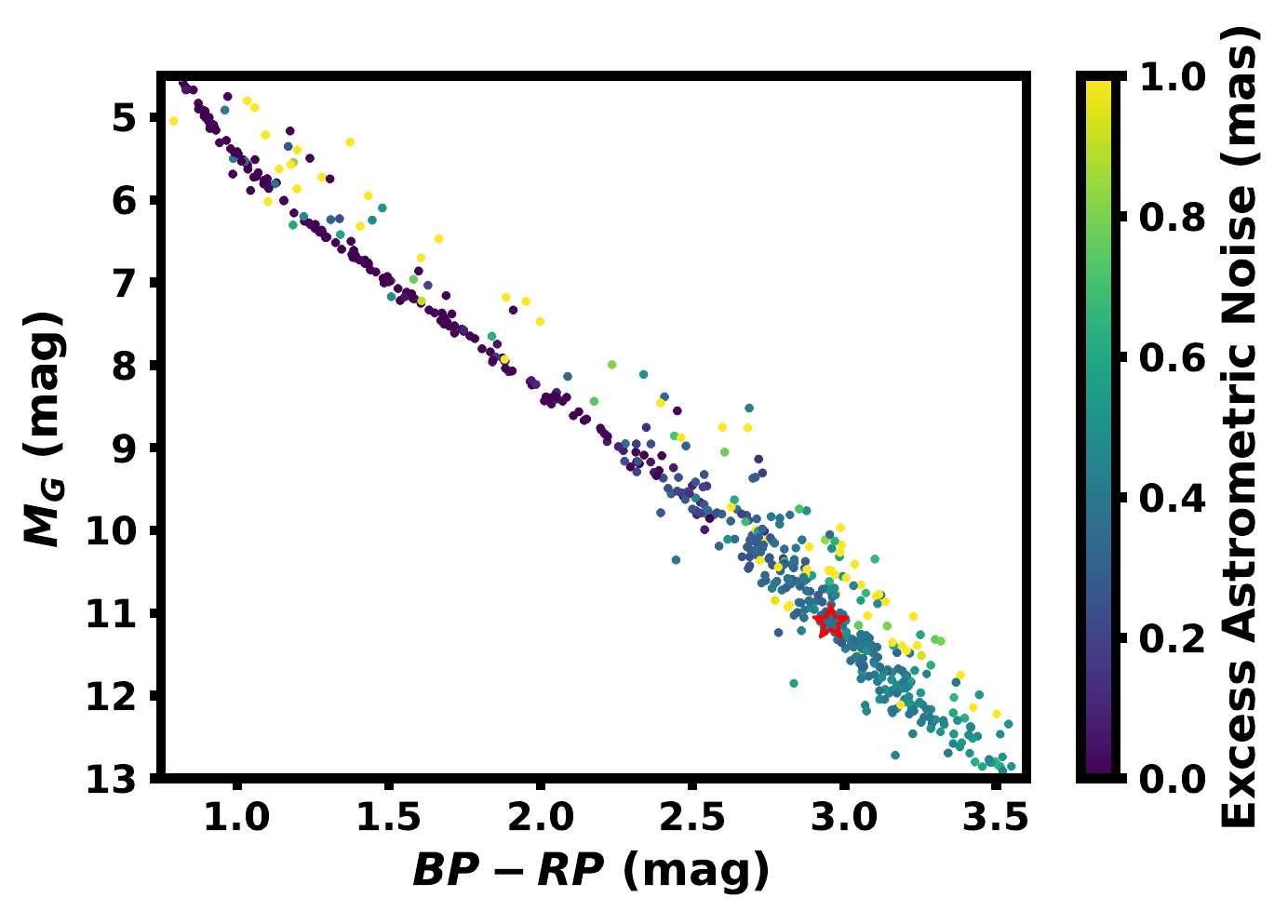}
    \hfill\includegraphics[width=0.47\textwidth]{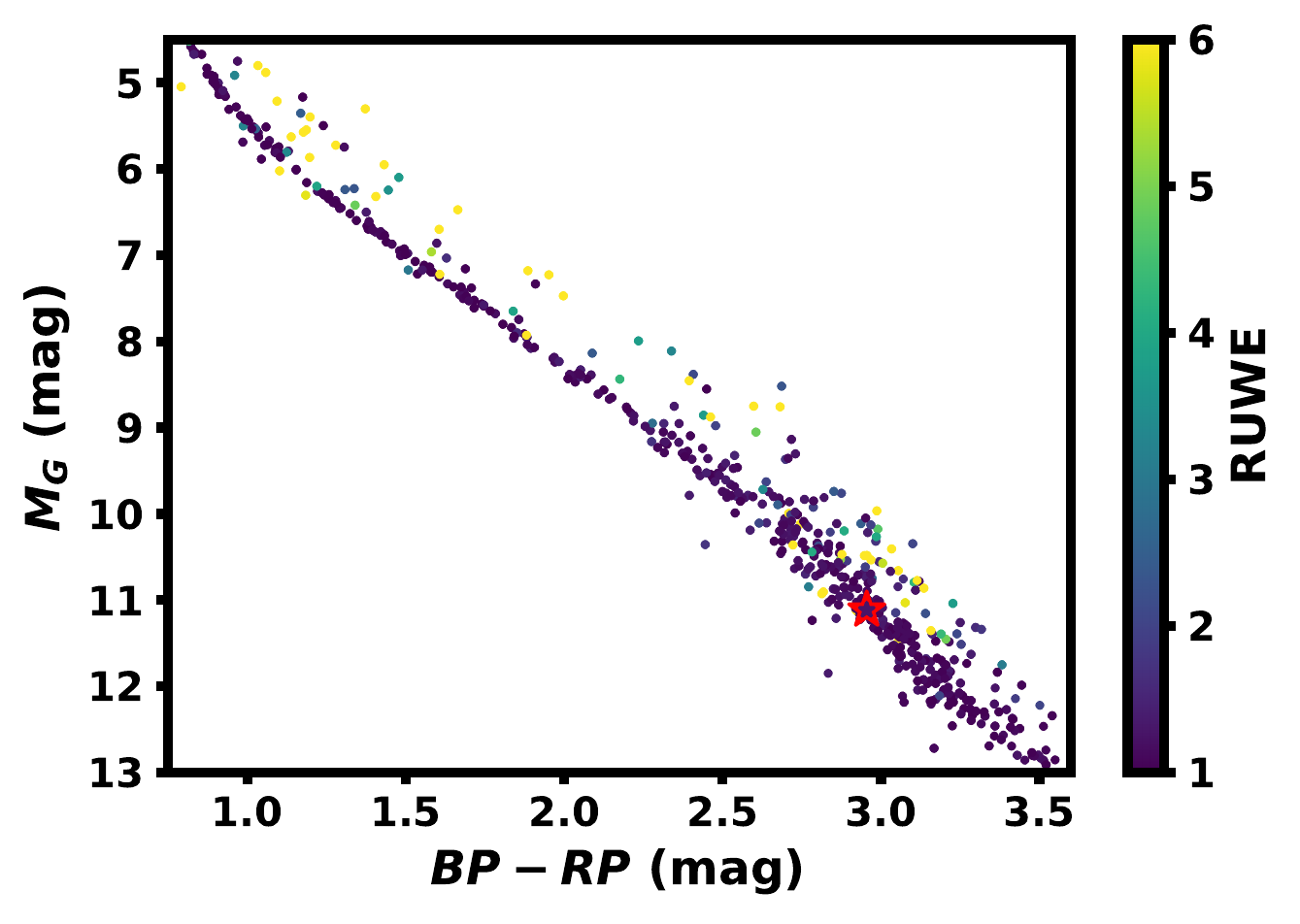}
\caption{Color-magnitude diagram of Hyades members identified in \citet{Rizzuto2017} with parallaxes and photometry drawn from \textit{Gaia} DR2. Points are color-coded by their excess astrometric noise (top) and RUWE (bottom) reported in DR2. Excess astrometric noise and RUWE can go beyond the color range shown, but color is truncated for clarity. K2-25 is outlined with a red star and filled following the same coloring as other stars. Systems on the binary sequence often show increased astrometric noise and a larger RUWE when compared to those on the single-star sequence. K2-25 is consistent with similar-color single stars using either metric. \label{fig:noise}}
\end{figure}

Unresolved binarity could also bias the derived stellar density, as it would lead to an overestimation in the $K_S$-band magnitude of the target used in the empirical relations. The DR2 measurements for K2-25 had significant excess astrometric noise (\texttt{astrometric\_excess\_noise}$=0.374$\,mas, \texttt{astrometric\_excess\_noise\_sig}=23$\sigma$), which is a sign of binarity \citep[e.g.,][]{2018RNAAS...2...20E, rizzuto2018zodiacal}. However, redder stars also show significant astrometric noise independent of binarity. The renormalized unit weight error (RUWE\footnote{\href{https://www.cosmos.esa.int/web/gaia/dr2-known-issues}{https://www.cosmos.esa.int/web/gaia/dr2-known-issues}}) accounts for this color effect, making it a more reliable indicator of binarity \citep[e.g.,][]{ziegler2018measuring}. As we show in Figure~\ref{fig:noise}, K2-25's RUWE value is consistent with a single-star.

We further explored the impact of binarity on our result by assuming K2-25 is an equal-mass binary. While the color-magnitude diagram position and existing imaging and radial velocities for K2-25 ruled out such a scenario, this test represents the most extreme case in terms of impact on the assigned stellar parameters. For this test, we recomputed $M_*$, $R_*$, and $\rho_*$ as in Section~\ref{sec:params}, but first subtracted half the flux from the $K_S$-band magnitude. This yielded $\rho_*=17\pm2 \rho_{\odot}$, which was still inconsistent with the $e=0$ transit-fit stellar density at 6$\sigma$ (see Figure~\ref{fig:dencomp}). The presence of an unresolved host would also impact the transit, but dilution primarily affects the transit depth, not the duration \citep{Kraus2016a, Mann2017b, Teske2018}.

We conclude that the large eccentricity derived for K2-25b's orbit from our transit light-curve analysis is not a consequence of complications in the data or input assumptions about the host star. 

\subsection{Comparison to Model Transmission Spectra}

\begin{figure*}[ht!]
    \plotone{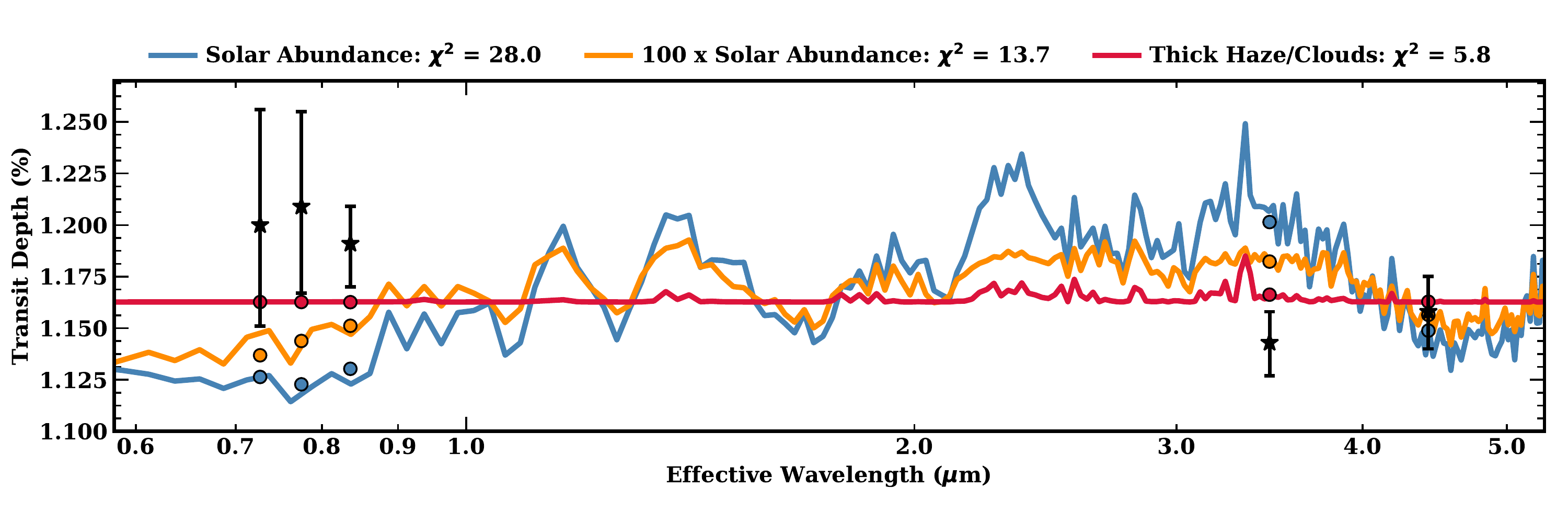}
    \caption{Transmission spectra of K2-25b from our data (black stars) compared to a solar abundance atmosphere (blue line), 100$\times$ solar abundance atmosphere (orange line), and a thick haze/cloud atmosphere (red line) model, assuming the planet's mass is 11.5$M_{\oplus}$ and the equilibrium temperature is 400 K. The filled circle points indicate the corresponding synthetic transit depth from convolving the spectrum with the appropriate filter profile. All models were normalized to give the best fit to the data. All calculations were done with high-resolution models; the models shown here are binned to a median width of $\sim$0.025$\mu$m.
    \label{fig:trans_spectra_9_400_11.5}}
\end{figure*}

We present the transmission spectrum of K2-25b in Figure~\ref{fig:trans_spectra_9_400_11.5}. Due to degeneracies between atmospheric parameters, the unknown planet mass, and the limits of broadband data, we were not able to fully probe the content of the atmosphere. Instead, our goal was to test if K2-25b's atmosphere is more consistent with a flat or featured spectrum. To this end, we compared our results to model spectra for three atmospheric scenarios: a solar abundance atmosphere, 100 $\times$ solar abundance atmosphere, and a cloudy atmosphere (flat transmission spectrum). 
The model spectra were generated by the publicly available \texttt{Exo$-$Transmit} open source code\footnote{\href{https://github.com/elizakempton/Exo_Transmit}{https://github.com/elizakempton/Exo\textunderscore Transmit}} \citep{kempton2017exo} with the included opacity data \citep{freedman2008line, freedman2014gaseous, lupu2014atmospheres}. As inputs, we set the radius of the planet to 3.45$R_{\oplus}$ and the radius of the star to 0.29$R_\odot$ (Table~\ref{tab:gfparam} and Section~\ref{sec:params}). We use the nominal setting for Rayleigh scattering, assumed equilibrium chemistry, and included condensation and removal via rainout of molecules. We varied the equilibrium temperature, surface gravity, and metallicity. We ran models using metallicities ([M/H]) of 1$\times$ solar and 100$\times$ solar. The equilibrium temperatures we tested were 300, 400, and 500 K, with 400 K being the orbit-averaged equilibrium temperature of the planet assuming an albedo of 0.3. Since the mass is unknown, we assigned a mass using the the mass-radius relations from
\cite{2016ApJ...825...19W}, which yielded $M_{P}= 13\pm2M_{\oplus}$. With the planet being young, we expect the planet to be less dense than its older counterparts, so we tested three surface gravity ($g$) values: 6, 9, and 12 m\,s$^{-2}$, which corresponds to planet masses of $\simeq$8, 11.5, and 15$M_{\oplus}$. For the thick cloud/haze model, we set the pressure at cloud top to 10 Pa. In total, we had 19 model spectra. 
To compare our data to the models, we convolved the spectrum with the relevant filter profile to create a synthetic transit depth corresponding to each effective wavelength (photon weighted mean wavelength). The effective wavelength factors in the widths of the broadband filters and was calculated using K2-25's spectrum and each filter's bandpass. The results of this calculation yielded effective wavelengths of 0.73$\mu$m (\textit{K2}), 0.77$\mu$m (LCOGT), 0.84$\mu$m (MEarth), 3.46$\mu$m (Channel 1), 4.43$\mu$m (Channel 2). We added a free parameter to allow each model spectra to shift in median depth, and varied it to minimize the $\chi^2$ when compared to our data (to allow for small deviations in $R_P/R_*$). The normalization parameters and $\chi^2$ values are listed in Table~\ref{tab:models_chi_square}. 

Our results disfavor a cloud-free atmosphere in chemical equilibrium assuming solar abundance ($>4\sigma$ confidence), regardless of the equilibrium temperature and surface gravity value. Overall, the measured transmission spectrum is consistent with a flat line ($\chi^2$ = 5.8, dof=4), which is evidence of a cloudy/hazy atmosphere and/or a high mean molecular weight atmosphere. 

\begin{deluxetable} {lcccc} 
\tabletypesize{\footnotesize}
\tablecaption{Normalization Factor for Atmospheric Models  \label{tab:models_chi_square}}
\tablecolumns{5}
\tablenum{6}
\tablewidth{0pt}
\tablehead{
\colhead{Model} &
\colhead{$T_{\rm{eq}}$} & 
\colhead{$g$} &
\colhead{Normalization} & 
\colhead{$\chi^2$} \\ 
\colhead{} &
\colhead{(K)} & 
\colhead{(m\,s$^{-2}$)} & 
\colhead{Factor} & 
\colhead{(dof=4)}
} 
\startdata
  & 300 & 6 & 1.015 & 25.3 \\ 
  & 300 & 9 & 0.992 & 16.8 \\ 
  & 300 & 12 & 0.980 & 13.2 \\ 
  & 400 & 6 & 1.049 & 46.1\\ 
Solar Abundance & 400 & 9 & 1.013 & 28.0 \\ 
  & 400 & 12 & 0.995 & 20.6 \\ 
  & 500 & 6 & 1.088 & 78.6\\ 
  & 500 & 9 & 1.037 & 45.0 \\ 
  & 500 & 12 & 1.013 & 31.4 \\ 
\hline
  & 300 & 6 & 1.005 & 12.9 \\ 
  & 300 & 9 & 0.986 & 10.0 \\ 
  & 300 & 12 & 0.977 & 8.7 \\ 
  & 400 & 6 & 1.029 & 19.0 \\ 
100 $\times$ Solar Abundance & 400 & 9 & 1.00 & 13.7 \\ 
  & 400 & 12 & 0.988 & 11.3 \\ 
  & 500 & 6 & 1.072 & 37.0 \\ 
  & 500 & 9 &  1.028 & 23.6 \\ 
  & 500 & 12 & 1.007 & 17.8 \\ 
\hline
\hline
Thick Haze/ Clouds & $\dots$ & $\dots$ & 0.951 & 5.8
 \enddata
\end{deluxetable}

\subsubsection{Impacts of Stellar Variability/Spots}  

The transit depths in the infrared data are statistically smaller than the transit depths from the optical data (Figure~\ref{fig:trans_spectra_9_400_11.5}). This may be in part, due to inhomogeneities on the surface of on K2-25 (e.g., spots and plages). Unocculted spots make the transit appear deeper (because the planet blocks a statistically brighter part of the star), while a planet crossing directly over a large spot can similarly make the transit shallower (and vice versa for plages or faculae). Since these surface features vary in intensity as a function of wavelength, their impact on the inferred transit depth can have a strong impact on the transmission spectrum \citep[e.g., ][]{ 2015ApJ...814...66K, rackham2017access}

Only one light curve (transit number 195 from \spitzer) showed morphology consistent with K2-25b occulting a large spot, and this is only considered a candidate crossing due to the PSF landing outside the ``sweet spot'' \citep[see][]{Kain2019}. Smaller spots under the transit chord would be harder to detect, but could be noticed as an increased scatter in the data or time-variable transit depth. However, transit photometry residuals are similar both during and outside the transit across all observations. Either the transit chord is relatively pristine, or the spots are too small and/or evenly distributed, resulting in an insignificant impact on our final transit parameters.

The effect of spots in the near-infrared was small due to the low spot contrast at longer wavelengths. We expected the overall impact on the transit depth to be $\ll$100 ppm (per transit) past 2\um\ based on analyses of similarly variable stars \citep{rackham2017access}. This is smaller than the measurement errors and hence, unlikely to change our results from \spitzer. 

The impact of spots on the optical data were similarly mitigated by averaging over many transits. Long-term monitoring of K2-25 from MEarth \citep[][]{Kain2019} indicated that the overall stellar variability changed from 1.5\% in the 2015-2016 observing season to $<$0.5\% in 2016-2017, with a moderate increase to 0.5\% in 2017-2018. Such a large change in variability is most likely driven by changes in the overall spot coverage fraction of K2-25, yet there is no correlation between the stellar rotational phase and the planet-to-star radius ratio. While spots are clearly present on K2-25, as can be seen in the large stellar variability, this may be a large number of semirandomly distributed solar-like spots rather than a single giant spot or spot groups.

The impact of spots on the transmission spectrum of K2-25b is dependent on how the spots are distributed with respect to the transit chord. Even if the spot pattern changed between transits, if the active regions are offset from the transit chord, there may be a statistical preference for spots to appear not in the path of the transit -- biasing the inferred transit depth at bluer wavelengths. To model how this impacts our conclusions about the transmission spectrum of K2-25b, we estimated the fraction of spots required to reconcile the optical and near-infrared transit depths (i.e., to achieve a flat transmission spectrum). Following \citet{Rackham2018}, the effect of unocculted spots on the observed transit depth is approximately:
\begin{equation}
    \delta_{\lambda,\rm{obs}} = \frac{\delta_{\lambda,\rm{true}}}{1-f_{spot}(1-\frac{F_{\lambda,\rm{star}}}{F_{\lambda,\rm{spot}}})},
\end{equation}
where $\delta_{\lambda,\rm{true}}$ and $ \delta_{\lambda,\rm{obs}}$ are the true and observed transit depths as a function of wavelength, $f_{spot}$ is the fractional spot coverage of the star assuming the transit chord is spot-free, and $F_{\lambda,\rm{spot}}$ and $F_{\lambda,\rm{star}}$ are the spectra of the spot and unspotted star, respectively. For this test, we ignored the effect of faculae/plagues, as these primarily increase the discrepancy between optical and NIR transit depths. 

To estimate $F_{\lambda,\rm{spot}}$ and $F_{\lambda,\rm{star}}$, we used BT-SETTL models \citep{Allard2013} assuming spots temperatures of 2800\,K or 3000\,K on a stellar surface of 3200\,K. We convolved each atmosphere model with the relevant filter profiles for each of the five observations (\kepler, LCOGT, MEarth, and the two \spitzer\ bands). We then fit for the best $f_{\rm{spot}}$ for an assumed set of $\delta_{\lambda,\rm{true}}$ values. To match the predictions of a flat transmission spectrum required spot coverage fractions of 11\% for 2800\,K spots and 17\% for spots of 3000\,K (1-2\% coverage is allowed at 2$\sigma$). For the observed transit depths to match the solar-composition model-predicted values required spot coverage fractions of 22\% and 36\%, with values of $f_{\rm{spot}}$ as small as 14\% and 24\% to be consistent with the model at 2$\sigma$. These $f_{\rm{spot}}$ estimates are relative to the spot fraction in the transit chord, so the true spot fractions required are likely larger.

While large spot coverage fractions ($>50\%$) have been observed in young stars \citep[e.g.,][]{Gully-Santiago2017}, these systems tend to show high overall stellar variability (10-30\%). The $f_{\rm{spot}}$ values needed to fit the solar-abundance model are hard to reconcile with the observed out-of-transit variability seen from K2-25 ($\lesssim$2\% in \ktwo\ data and $<1\%$ in MEarth).
Large spots would likely produce variations in transit depth between transits  (as the planet crosses different regions of the star or the spot pattern changes), yet transit depths are consistent over multiple years and show no significant red noise in the fit residuals (see Figure~\ref{fig:lc_all}). The impact of large spots would be small at \spitzer\ wavelengths (and we selected PMAP corrections in part because of consistency between transits), but easily visible in both \ktwo\ and MEarth data (optical wavelength range). Neither show significant red noise or obvious spot crossings. A large number of small spots with semi-even distribution over the star could produce a small variability profile with a large spot coverage fraction, although this decreases the probability that the transit chord is pristine and makes the smooth sinusoidal variation outside of the transit harder to fit. More realistic simulations of spot fractions suggested that during the low variability season (0.5\% flux variation in MEarth), spot coverage fractions correspond to  1\%-10\% \citep{Rackham2018}.
 
The best-fit spot coverage fractions assuming the solar-abundance model also provide a relatively poor fit to the data, as spots alone cannot explain the consistency between the depths in the two \spitzer\ bands (See Figure~\ref{fig:trans_spectra_9_400_11.5}). We conclude that while spots likely do have an impact on our overall transmission spectrum, they are unlikely to change the results enough to be consistent with the predictions from the solar-abundance atmosphere model.

\section{Summary and Conclusions}\label{sec:summary}

To constrain K2-25b's dynamical history, refine the planet's properties, and explore its transmission spectrum, we combined transit observations from ground-based (LCOGT and MEarth) and space-based (\textit{K2} and \textit{Spitzer}) facilities, totaling 44 transits over $>2$ yr. Our analysis of these data included comparing corrections for \textit{Spitzer}'s large intra-pixel sensitivity variations using three different techniques: PLD, PMAP, and NNBR--all of which yielded consistent transit parameters.

PMAP corrections performed best based on consistency between transit depths between transits and minimizing red noise, in apparent contradiction to previous publications \citep[e.g.,][]{Ingalls2016}. However, we caution against interpreting these results as evidence of PMAP being superior. All three methods yielded RMS levels near the white-noise limit, likely because the PSF was stable (Section~\ref{sec:spitzer}) and the transit duration of K2-25b is much shorter than most transiting systems. As PMAP is most sensitive to centroid variations and the PSF landing off the sweet spot, stability yielded a higher-than-typical performance for PMAP. This also makes K2-25b a poor case to draw general conclusions, especially for more typical ($\gg2$hr) observing windows. We encourage others to inspect more detailed tests done on more typical systems \citep[e.g.,][]{Ballard2014, Ingalls2016, 2016ApJ...820...86M, 2017AJ....153...22K, 2017PASP..129a4001S}

To constrain the parameters of K2-25b, we fit the extracted light curves at each mean wavelength (0.64, 0.76, 0.83, 3.56, and 4.50$\mu$m), as well as all data together simultaneously within an MCMC framework. The combined dataset demonstrated that K2-25b's orbit is significantly eccentric (> 0.20), independent of the dataset or wavelength of the observations. This result is consistent with the findings of \cite{vanEylen_eccentricity2019}, that single transiting planets have a higher eccentricity compared to multiple-transit systems, and hints this bimodal distribution of eccentricities seen in older planets arises earlier in the planet's history or during formation, although a larger sample of planets (including young multitransiting systems) will be needed to test this.

This high eccentricity is suggestive that this planet has a complex dynamical history \citep{juric2008dynamical, davies2013long} and motivates further searches for stellar companions or additional planets \citep[e.g.,][]{Kain2019}. Existing radial velocity and adaptive optics imaging of the host taken as part of the discovery paper, as well as \textit{Gaia} imaging \citep{rizzuto2018zodiacal, ziegler2018measuring} ruled out the tightest and widest companions. Further radial velocity observations of K2-25 would be invaluable to confirm the high eccentricity while simultaneously searching for additional companions that may impact the system's evolution. 

The observed transmission spectrum from our transit observations disfavors a solar-abundance atmosphere at $>4\sigma$ for any reasonable planet mass (8, 11.5, 15$M_{\oplus}$) and equilibrium temperature (300, 400, 500 K). The transmission spectrum of K2-25b is consistent with being flat, suggesting that the planet has a featureless (or weakly featured) transmission spectrum. This result follows the findings of \cite{Crossfield2017} that Neptune atmospheres cooler than $T_{\rm{eq}}$ = 800 K tend to be featureless. Our results are also consistent with the predictions of \citet{Wang2019}, which suggest that young planets may have outflowing atmospheres with small dust grains that result in flat transmission spectra and inflated radii. 

Unocculted spots during transits may be systematically biasing our fits of optical transit photometry to larger depths (by making the transited region brighter than the stellar average), with a relatively smaller impact on the \spitzer\ data. This effect can explain the difference between our optical and NIR transit depths given a flat transmission spectrum, but the spot fraction required to explain the solar-abundance model (22-36\%) are unlikely given the 0.5\%-2\% stellar variability seen in the out-of-transit data and assumption of a pristine transit chord, and multiyear stability of the transit depth in the optical. More detailed modeling or additional data will be required to make more definitive statements about the mean molecular weight and/or presence of clouds/hazes in the atmosphere of K2-25b. Additional transits at $JHK$ bands from \textit{Hubble Space Telescope}/WFC3 or broadband data from the ground \citep[e.g. LUCI on LBT;][]{2017AJ....154..242B} would also be less impacted by spots than the optical data used here, and when combined with the \spitzer\ transits could confirm our findings. 

\acknowledgements

The authors thank the anonymous referee for comments that helped improved the manuscript. The authors would also like to thank Fei Dai for insightful conversations as the manuscript was being finalized. In addition, the authors wish to acknowledge Wally and Bandit for their emotional support during the writing of this manuscript, as well as their unwavering dedication to the sciences. 

P.C.T was supported by the TAURUS Scholars Program, which is funded in part by the University of Texas at Austin Department of Astronomy Board of Visitors and Cox Fund Endowment. The 2017 program was also supported through the generous donations of the public in the Fall 2016 HornRaiser campaign. TAURUS also receives support from NASA, NRAO and NSF.

This work was made possible by a grant from the {\it K2} Guest Observer Program (80NSSC19K0097) and support for the {\it Spitzer} observations through NASA's Astrophysics Data Analysis Program (80NSSC19K0583).

This paper includes data collected by the K2 mission. Funding for the K2 mission is provided by the NASA Science Mission directorate. This work makes use of observations made in the LCOGT network. This work is based on observations made with the \spitzer\ \textit{Space Telescope}, which is operated by the Jet Propulsion Laboratory, California Institute of Technology under a contract with NASA. Support for this work was provided by NASA through an award issued by JPL/Caltech. The MEarth Team gratefully acknowledges funding from the David and Lucille Packard Fellowship for Science and Engineering (awarded to D.C.). This material is based upon work supported by the National Science Foundation under grants AST-0807690, AST-1109468, AST-1004488 (Alan T. Waterman Award), and AST-1616624. This publication was made possible through the support of a grant from the John Templeton Foundation. The opinions expressed in this publication are those of the authors and do not necessarily reflect the views of the John Templeton Foundation.  

\software{\texttt{misttborn.py}, \textit{emcee} \citep{Foreman-Mackey2013}, \textit{batman} \citep{Kreidberg2015}, matplotlib \citep{hunter2007matplotlib}, \texttt{corner.py} \citep{foreman2016corner}, Exo-Transmit \citep{kempton2017exo}}

\facilities{\spitzer\ (IRAC), \ktwo\ , \mearth\ , LCOGT}

\clearpage
\bibliography{bib.bib}

\end{document}